%% file: 0-main.tex
\documentclass[10pt,conference]{IEEEtran} 
\IEEEoverridecommandlockouts

\usepackage{balance}
\usepackage{verbatim}
\usepackage{subfigure}
\usepackage{enumitem}
\usepackage{diagbox}
\usepackage{booktabs}
\usepackage[ruled,vlined,linesnumbered,boxed,commentsnumbered]{algorithm2e}  
\usepackage{algpseudocode}
\usepackage{color}
\usepackage{url}
\usepackage{multirow}
\usepackage{threeparttable}

\usepackage[framemethod=TikZ]{mdframed}
\usepackage{lipsum}
\usepackage{amsmath}
\usepackage{tipa}
\usepackage{xcolor}
\usepackage{framed}
\usepackage{soul}

\SetCommentSty{mycommfont}

\usepackage{stmaryrd}
\usepackage{wasysym}
\usepackage{cite}
\usepackage{tcolorbox}
\newcounter{num}

\usepackage{tikz}
\usepackage{pgfplots}
\newcommand*\circled[1]{\tikz[baseline=(char.base)]{
    \node[shape=circle,draw,inner sep=0.5pt] (char) {\small#1};}}

\newcommand{\qingchao}[1]{{\color{teal}[Qingchao: #1]}}

\newcommand{\tool}{\textsc{Opera}}
\newcommand{\tech}{\textsc{Opera}}

\newcommand{\TotalBugsNum}{$170$}
\newcommand{\ConfirmedBugsNum}{$90$}
\newcommand{\ConfirmedFrontBugsNum}{$87$}

\newcommand{\TotalCrashBugs}{101}
\newcommand{\TotalConfirmedCrashBugs}{59}
\newcommand{\TotalWrongBugs}{69}
\newcommand{\TotalConfirmedWrongBugs}{31}

\newcommand{\TVMBugsNum}{$79$}

\newcommand{\TRTBugsNum}{$43$}

\newcommand{\OVBugsNum}{$48$}

\newcommand{\NNSmithBugsNum}{$18$}
\newcommand{\NNSmithTVMBugsNum}{$11$}
\newcommand{\NNSmithTRTBugsNum}{$2$}
\newcommand{\NNSmithOVBugsNum}{$5$}

\newcommand{\COMETTVMBugsNum}{$6$}
\newcommand{\COMETTRTBugsNum}{$0$}
\newcommand{\COMETOVBugsNum}{$4$}

\newcommand{\AvgAPFD}{$0.898$}

\newcommand{\APFDRandomImprove}{13.1\%}
\newcommand{\APFDFastImprove}{11.9\%}
\newcommand{\APFDTotalImprove}{47.4\%}
\newcommand{\APFDAdditionImprove}{37.2\%}

\newcommand{\ie}{\hbox{\emph{i.e.}}\xspace}

\newcommand{\viz}

\makeatletter
\newcommand{\linebreakand}{%
  \end{@IEEEauthorhalign}
  \hfill\mbox{}\par
  \mbox{}\hfill\begin{@IEEEauthorhalign}
}
\makeatother

\begin{document}
\title{A Tale of Two DL Cities: \\When Library Tests Meet Compiler}

\author{
\IEEEauthorblockN{Qingchao Shen}
\IEEEauthorblockA{\textit{College of Intelligence and }\\
\textit{Computing, Tianjin University}\\
Tianjin, China \\
qingchao@tju.edu.cn}
\and
\IEEEauthorblockN{Yongqiang Tian}
% \IEEEauthorblockA{\textit{Department of Computer Science }\\
% \textit{and Engineering, The Hong Kong }\\
% \textit{University of Science and Technology}\\
\IEEEauthorblockA{
\textit{The Hong Kong University}\\
\textit{ of Science and Technology}\\
Hong Kong, China \\
yqtian@ust.hk}
\and
\IEEEauthorblockN{Haoyang Ma}
\IEEEauthorblockA{
% \textit{Department of Computer Science }\\
% \textit{and Engineering, The Hong Kong }\\
% \textit{University of Science and Technology}\\
\textit{The Hong Kong University}\\
\textit{ of Science and Technology}\\
Hong Kong, China \\
haoyang.ma@connect.ust.hk}
\and
\IEEEauthorblockN{Junjie Chen\dag}
\thanks{\dag Junjie Chen is the corresponding author}
\IEEEauthorblockA{\textit{College of Intelligence and }\\
\textit{Computing, Tianjin University}\\
Tianjin, China \\
junjiechen@tju.edu.cn}
\and
\IEEEauthorblockN{Lili Huang}
\IEEEauthorblockA{\textit{College of Intelligence and }\\
\textit{Computing, Tianjin University}\\
Tianjin, China \\
huangll@tju.edu.cn}
\and
\IEEEauthorblockN{Ruifeng Fu}
\IEEEauthorblockA{\textit{College of Intelligence and }\\
\textit{Computing, Tianjin University}\\
Tianjin, China \\
frf2000@tju.edu.cn}
\and
% \linebreakand
\IEEEauthorblockN{Shing-Chi Cheung}
\IEEEauthorblockA{
% \textit{Department of Computer Science }\\
% \textit{and Engineering, The Hong Kong }\\
% \textit{University of Science and Technology}\\
\textit{The Hong Kong University}\\
\textit{ of Science and Technology}\\
Hong Kong, China \\
scc@cse.ust.hk}
\and
\IEEEauthorblockN{Zan Wang}
\IEEEauthorblockA{\textit{College of Intelligence and }\\
\textit{Computing, Tianjin University}\\
Tianjin, China \\
wangzan@tju.edu.cn}
}

\maketitle

\pagestyle{plain}
\begin{abstract}
Deep Learning (DL) compilers typically load a DL model and optimize it with intermediate representation. 
Existing DL compiler testing techniques mainly focus on model optimization stages, but rarely explore bug detection at the model loading stage.
Effectively testing the model loading stage requires covering diverse usages of each DL operator from various DL libraries, which shares a common objective with DL library testing, indicating that the embedded knowledge in DL library tests 
% could potentially be 
is beneficial for testing the model loading stage of DL compilers.
With this idea, we propose \tool{} to migrate the knowledge embedded in DL library tests to test the model loading stage.
\tool{} constructs diverse tests from various tests for DL libraries (including the tests documented in DL libraries and those generated by recent fuzzers).
In total, we considered three sources of tests in DL libraries for migration.
In addition, it incorporates a diversity-based test prioritization strategy to migrate and execute those tests that are more likely to detect diverse bugs earlier.
% Thus, we conducted the first empirical study to investigate the effectiveness and efficiency of migrating the knowledge embedded in DL library tests to test the model loading stage. 
% To support the conduct of this study, we develop a technique, called \tool{}, consisting of test migration (regarding effectiveness investigation) and test prioritization (regarding efficiency investigation).
We then used eight frontends from three DL compilers (e.g., TVM, TensorRT, and OpenVINO) for evaluation.
% The migrated tests with the aid of 
\tool{} detected \TotalBugsNum{} previously unknown bugs in total, \ConfirmedBugsNum{} of which have been confirmed/fixed by developers,
demonstrating the effectiveness of such the migration-based idea.
The test prioritization strategy in \tool{} improves testing efficiency with migrated tests by \APFDFastImprove{}$\sim$\APFDTotalImprove{} on average compared to general test prioritization strategies.
% Finally, we obtained 7 major findings and provided a set of guidelines for future work from this study.
\end{abstract}


%    <concept>
%        <concept_id>10011007.10011074.10011099.10011102.10011103</concept_id>
%        <concept_desc>Software and its engineering~Software testing and debugging</concept_desc>
%        <concept_significance>500</concept_significance>
%        </concept>
%  </ccs2012>

% \ccsdesc[500]{Software and its engineering~Software testing and debugging}


\input{1-intro}

\input{2-background}

\input{3-methodology}

\input{4-experiment-setup}

\input{5-evaluation}

\input{5.2-discussion}

\input{6-related}

\input{7-conclusion}
\input{8acknowledge}

% \begin{acks}

% \end{acks}

\balance
\bibliographystyle{IEEEtran}
\bibliography{ref}

\end{document}

%% file: 1-intro.tex
\section{Introduction}
\label{sec:intro}
Deep Learning (DL) compilers (e.g., TVM~\cite{tvm}, TensorRT~\cite{TensorRT}, and OpenVINO~\cite{OpenVINO}) are widely utilized to optimize the performance of DL models for deployment on various hardware devices.
The compilation process of a DL model typically involves three main stages~\cite{shen2021comprehensive}:
\circled{1} Loading the DL model, prepared under a specific DL library (e.g., PyTorch~\cite{pytorch} or Keras~\cite{keras}), into its equivalent high-level intermediate representation (IR);
\circled{2} Performing hardware-independent optimizations on the high-level IR;
\circled{3} Lowering the high-level IR into the low-level IR and conducting hardware-specific optimizations to generate code targeting specific hardware.

Similar to traditional compilers~\cite{sun2016toward, zhou2021empirical, jvmlanguagebugstudy, pyinterpreterstudy,chen2023compiler,wu2023jitfuzz,chen2020enhanced}, DL compilers also contain bugs, which can compromise the reliability of both the compilers themselves and the models they produce.
As reported~\cite{shen2021comprehensive}, each stage of DL compilers contains a significant number of bugs, underscoring the need for comprehensive testing to ensure the quality of DL compilers. 
However, existing DL compiler testing techniques primarily focus on the two optimization stages (\circled{2} and \circled{3}) mentioned earlier, neglecting the model loading stage (\circled{1}). 
Specifically, recent DL compiler testing techniques (such as HirGen~\cite{ma2022hirfuzz}, Tzer~\cite{tzer}, and TVMFuzz~\cite{TVMFuzz}) mainly construct tests at either high-level or low-level IRs, bypassing the model loading stage.

NNSmith, a state-of-the-art grammar-based technique~\cite{NNSmith}, is designed to construct DL models for testing various stages of DL compilers. 
However, it primarily focuses on stress testing for optimizations by generating complicated models. 
In contrast, the model loading stage involves converting each operator in a model into its equivalent high-level IR individually. 
Therefore, effectively testing the model loading stage requires covering diverse usages of each DL operator from various DL libraries, rather than focusing on complex dependencies among operators. 
Furthermore, NNSmith is limited to constructing DL models solely under the ONNX library and supports a limited number of operators, making it ineffective for testing the model loading stage.

Intuitively, manually developing test generation tools that follow the corresponding grammars may meet the test requirements for the model loading stage. 
However, this method can be labor-intensive and error-prone due to the extensive number of DL libraries and their supported operators. 
Additionally, operators often involve numerous parameters, leading to complex constraints that further aggregate the difficulty of developing such tools. 
This highlights the need for alternative and lightweight methods to address this challenging task.

By analyzing source code, test cases, and bugs of DL compilers, we found that:
(1) Testing the model loading stage of DL compilers is related to testing DL libraries. 
Specifically, DL compilers typically accept DL models composed of operators supported by specific DL libraries as inputs.
Both the testing of the model loading stage and DL libraries share a common objective, which is to ensure the correctness of operators under various usages.
While there may not be complete overlap between the corner usages of each operator for DL compilers and DL libraries, the embedded knowledge in DL library tests could potentially be beneficial for testing the model loading stage of DL compilers.
(2) A few tests in DL compilers are designed with inspiration from tests documented in the ONNX library, as indicated by the comments accompanying these tests. 
This suggests the feasibility of leveraging the knowledge embedded in DL library tests to enhance the testing of the model loading stage to some extent.

However, due to the separate development of communities for DL compiler testing and DL library testing, there has been no systematic study to investigate the feasibility of migrating the knowledge embedded in DL library tests for testing the model loading stage. 
Hence, we performed the first exploration on the potential of the migration-based idea.
% \jj{why suddenly falling into test migration?--changed}
Specifically, we design a migration-based technique, called \tool{} (\underline{OPER}ator \underline{A}dapter), to test DL compilers (especially the model loading stage) by considering three sources of tests
% \jj{what's the difference between test inputs and tests?}
in DL libraries for migration, i.e., tests documented in DL libraries and tests generated by two recent fuzzers (DocTer~\cite{docter} and DeepREL~\cite{deepREL}).

In fact, the direct adoption of most DL library tests for testing the model loading stage of DL compilers is not feasible due to differences in their input formats.
Specifically, while DL compiler tests rely on pre-constructed DL models, DL library tests typically involve subtasks (e.g., gradient calculation and model design) in model construction, many of which cannot be represented in the form of DL models.
To address this challenge, \tool{} first extracts instances of DL operators from each DL library test via code instrumentation and then packages each operator instance to a model (as a migrated test for DL compilers) with the aid of model generation templates for different DL libraries.
Each operator instance represents a specific usage of the operator, encompassing an operator API and its corresponding parameter settings.

Another challenge that \tool{} encounters is the significant cost consideration.
% Assuming positive conclusions are drawn for the first RQ, the practicality of such a migration-based idea may still be hindered by significant cost considerations. 
This is primarily due to two factors: 
(1) the substantial volume of migrated tests originating from various migration sources, and 
(2) the frequent migration and execution of tests resulting from the frequent evolution of both DL libraries and DL compilers. 
% \textbf{This motivates the formulation of the second RQ, which investigates the efficiency improvement of this migration-based idea.}
To address this challenge, the component of test prioritization is designed in \tool{}, which prioritizes the migration and execution of tests that are more likely to uncover a diverse range of bugs in the model loading stage. 
After prioritization, more bugs can be detected within any given time budgets, thereby enhancing the overall test efficiency. 
The test prioritization component takes into account the diversity of operator instances.

In this work, 
% we consider three sources of tests in DL libraries for migration, i.e., tests documented in DL libraries and tests generated by two recent fuzzers (DocTer~\cite{docter} and DeepREL~\cite{deepREL}).
we applied 
% these migrated tests with 
\tool{} to test three popular DL compilers (i.e., TVM~\cite{tvm}, TensorRT~\cite{TensorRT}, and OpenVINO~\cite{OpenVINO}).
To balance evaluation cost and conclusion generality, for each compiler, we chose several popular DL libraries, i.e., PyTorch, Keras, and ONNX, from its supported frontends (responsible for model loading).
In total, our study covered eight frontends across three DL compilers.
In total, \tool{} detects \TotalBugsNum{} previously unknown bugs by migrating the knowledge embedded in DL library tests, \ConfirmedBugsNum{} of which have been confirmed/fixed, while the state-of-the-art grammar-based DL compiler testing technique (NNSmith) detects only \NNSmithBugsNum{} bugs within the same time budget.
The results demonstrate the effectiveness of such a migration-based idea for testing the model loading stage of DL compilers.
% (answering RQ1).
Furthermore, the test efficiency can be largely improved with the test prioritization component in \tool{}.
On average across eight subjects, it improves \tech{} without special test prioritization by \APFDRandomImprove{} and improves \tool{} incorporating the widely-used test prioritization strategies in general software testing by \APFDFastImprove{}$\sim$\APFDTotalImprove{} in terms of APFD (Average Percentage of Faults Detected)~\cite{apfd}.
% (answering RQ2).
% More findings and implications from our empirical study will be summarized in Section~\ref{sec:evaluation}.

This work makes the following major contributions:
\begin{itemize}[topsep=0pt, leftmargin=*]
    \item We introduced the idea of migrating knowledge from DL library tests to enhance the testing of the model loading stage in DL compilers.

    \item We designed a migration-based technique (\tool{}), which integrates various migration sources (i.e., tests documented in DL libraries and those generated by recent fuzzers), along with diversity-based test prioritization.
    
    \item We conducted an extensive study to evaluate \tool{} across eight frontends from three DL compilers, leading to the efficient detection of \TotalBugsNum{} previously unknown bugs.
    
    \item We released the \tool{} implementation and all experimental data for replication and future use, accessible at: \textbf{\url{https://github.com/ShenQingchao/OPERA}}.
\end{itemize}

%% file: 2-background.tex
% \vspace{-1mm}
% \section{Preliminaries}  
\section{Motivation}
\label{sec:background}

% \subsection{DL Compilers}
% \begin{figure}
%     \centering
%     \includegraphics[width=\linewidth]{figures/dlcompiler.pdf}
%     \vspace{-6mm}
%     \caption{The architecture of DL compilers}
%     \label{fig:dl_compiler}
% \end{figure}

% \jj{R2: this paper does not provide enough background on DNNs and their operators.}
% \jj{We may add the format of DL models and operators in Figure 1.}
% \qingchao{done, plz recheck.}
% DL compilers convert a DL model into an executable code on target devices.
% Figure~\ref{fig:dl_compiler} shows the architecture of DL compilers.
% The compilation process in DL compilers is divided into three main stages.
% according to their respective responsibilities. 
At the model loading stage, DL compilers take as input DL models built from various DL libraries,
% (\jj{add some background of DL models and operators, such as introducing the specific constructs and file formats}\qingchao{done!}), 
e.g., PyTorch, and convert them into a unified high-level IR. 
The model constructed by a specific DL library is a computational graph with diverse DL operators.
% The behavior of these operators can be customized by input parameters.
% A DL model is a directed acyclic graph with nodes representing operators (i.e., computational functions for tensors) and edges representing data flow.
This high-level IR, known as graph-level IR, helps hide the differences in DL models from various DL libraries, simplifying optimization execution. 
Each operator in the DL model is converted into semantically equivalent one or more IR expressions. 
For example, a \texttt{Conv2D} operator in a Keras model, along with all parameter settings (e.g., filters), is converted to \texttt{nn.conv2d} in the high-level IR of TVM.
% As the second stage of compilation, 
% \textit{high-level optimizations}
% perform
% \scc{In both abstract and introduction, we refer to this stage as an optimization stage. Are transformation stage referring to the optimization stage?}\qingchao{yes, the figure is changed}
% hardware-independent optimizations (\eg, operator fusion) 
% on the high-level IR 
% to reduce redundancy and improve efficiency.
% Third, at the \textit{low-level optimization} stage, the high-level IR is transformed into low-level IR, during which hardware-specific optimizations 
% (\eg, memory latency hiding) 
% are performed. 
% Finally, DL compilers generate executable code on specified target hardware from low-level IR.
% \victor{at least we should show the words ``high-level Optimization''
% and ``Lowe-level Optimization'' in this paragraph,
% since we are introducing the three stages of DL compilers, not only the loading stages.}\qingchao{changed, plz recheck}

\begin{figure}[t]
    % \centering  
    \subfigure[{\tt Conv2DTranspose} definition \newline in Keras]{
        \includegraphics[width=0.47\linewidth]{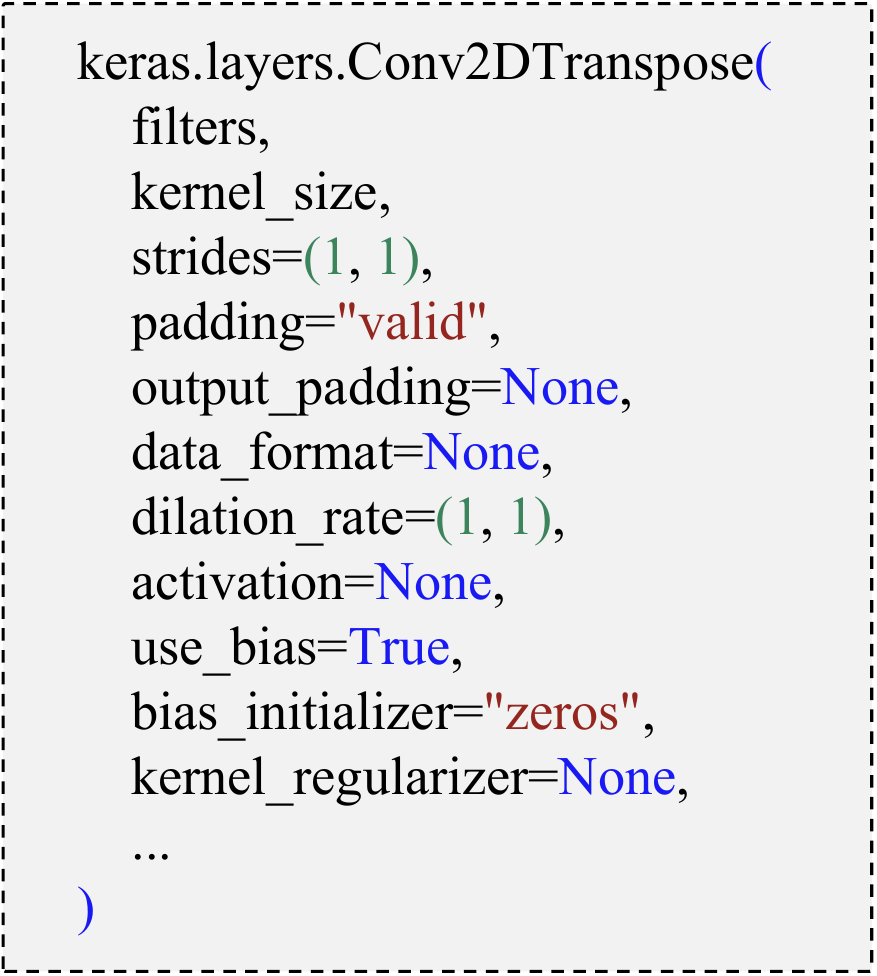} 
        \label{fig:motivation_example_api}
    }
    \hspace{-2mm}
    % \hfill
    \hfill
    \subfigure[A test for {\tt Conv2DTranspose} in the test suite of Keras]{
        \includegraphics[width=0.475\linewidth]{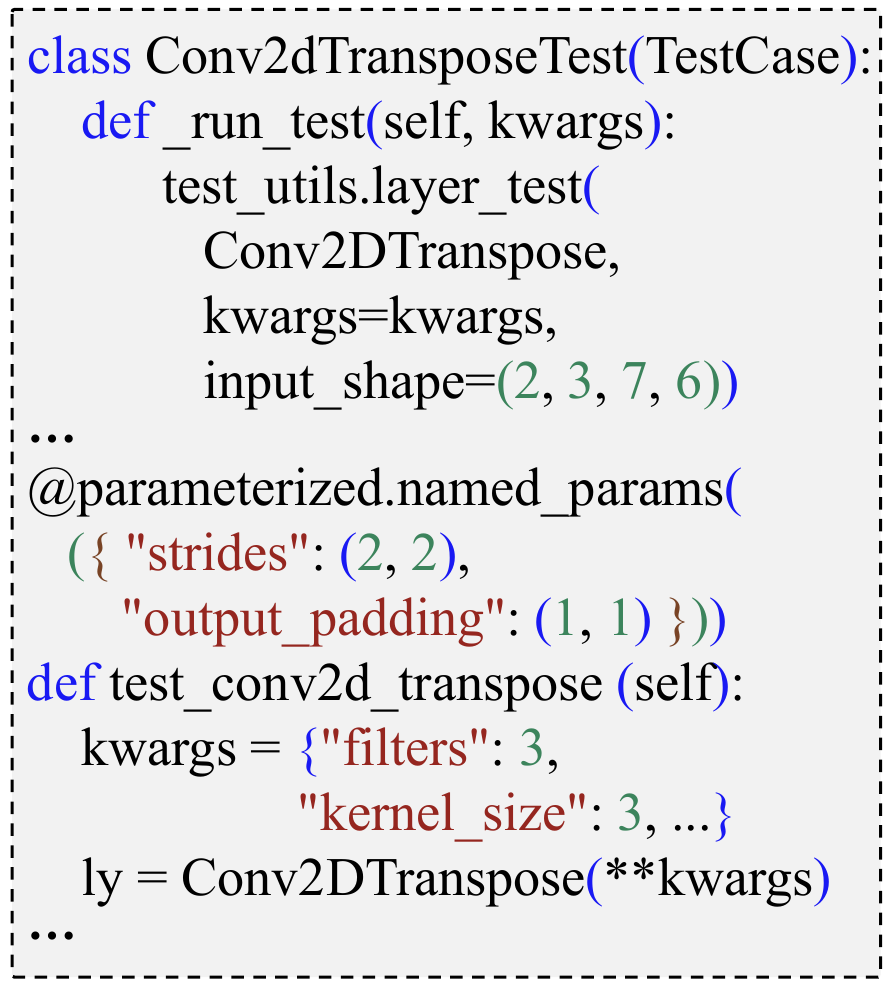}
        \label{fig:keras_test}
    }
    % \subfigure[The extracted operator instance for Conv2DTranspose]{
    %     \includegraphics[width=\linewidth]{figures/operator_instance.pdf}
    %     \label{fig:operator_instance}
    % }
    % \vspace{-2mm}
    \caption{A motivating example with {\tt Conv2DTranspose}}
    \label{fig:motivation_example} 
\end{figure}

% \subsection{A Motivating Example}
% \victor{I revised Motivating Example, please take a look}
% The model constructed by a specific DL library is a computational graph with diverse DL operators.
The behavior of operators can be customized by input parameters.
For example, Figure~\ref{fig:motivation_example_api} shows the definition of the \texttt{Conv2DTranspose} operator in Keras.
It includes two required parameters, \texttt{filters} and \texttt{kernel\_size}, and 14 optional parameters (e.g., strides) with default values.
The value selection of each parameter may affect the calculation result of a model involving the operator.
To guarantee the correctness of a DL library, numerous tests containing diverse operator instances with various parameter values were constructed for DL library testing~\cite{docter}.
For example, to test the correctness of the \texttt{Conv2DTranspose} operator, Keras developers prepared 84 tests.

In the model loading process, DL compilers need the ability to handle various usages of these operators, including all the combinations of parameters, for correct transformation from DL operators to high-level IR.
Hence, the model loading stage of DL compilers actually shares a similar test objective with the tests for these operators in DL libraries.
% \qingchao{\textbf{DL library vs DL compilers testing:}}
% \qingchao{
% Both of them aim to verify that the system under test properly manages each operator instance\jj{what's the purpose of this sentence?}.
This motivates 
% the migration of the tests for DL libraries to test the model loading stage.
% }
% This motivates the systematic study on the feasibility and effectiveness 
the idea of migrating knowledge embedded in DL library tests to test the model loading stage.

% \begin{figure}
%     \centering
%     \includegraphics[width=\linewidth]{figures/keras_test.png}
%     \caption{The test case \textit{Conv2DTransposeTest} in the equipped test suite of Keras}
%     \label{fig:keras_test}
% \end{figure}

\begin{figure}
    \centering
    \includegraphics[width=\linewidth]{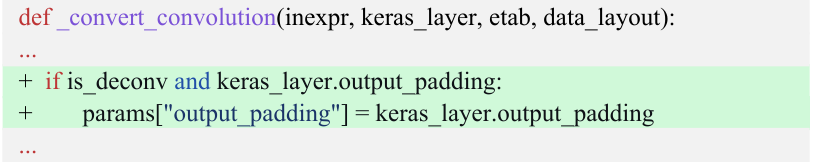}
    % \vspace{-6mm}
    \caption{Patch for a real bug on \texttt{Conv2DTranspose} in TVM}
    \label{fig:bug_motivation}
    % \vspace{-3mm}
\end{figure}

Figure~\ref{fig:bug_motivation} shows a real bug~\cite{bug_motivation}
% \footnote{https://github.com/apache/tvm/pull/15060} 
% \victor{I think use a citation instead of footnote is better here}
of \texttt{Conv2DTranspose} in TVM triggered by a test migrated from Keras testing.
TVM overlooks the parameter \texttt{output\_padding} when converting the operator \texttt{Conv2DTranspose}, which leads to incorrect output shape when the \texttt{out\_padding} is not set to the default value (i.e., {\tt None}).
% This test input invokes this API with a non-default value for the parameter \texttt{output\_padding}, which 
Before this bug was reported, the developer-provided tests of TVM only consisted of two operator instances to test the conversion of \texttt{Conv2DTranspose} from the DL operator to high-level IR. 
Moreover, in the two instances, all the optional parameters use the default values and thus are ineffective in detecting the bugs that require other, non-default parameter settings.

The tests migrated from DL library testing can help detect these cases.
The test with non-default \texttt{out\_padding} is absent in the tests of DL compilers, but available in the tests of Keras.
In Figure~\ref{fig:keras_test}, the test \texttt{Conv2dTransposeTest} from Keras, which assigns {\tt (1, 1)} to \texttt{out\_padding}, matches the bug-triggering condition for this bug, and thus can help reveal it.
% , resulting in wrong inference results.
% \victor{what is the symptom of this bug?}\qingchao{added}.
By migrating the Keras test to the test in the input format for TVM with this operator instance, \tool{} successfully detected this previously unknown bug.
After this bug was reported, it was fixed by adding the analysis on the parameter \texttt{output\_padding} in the \texttt{\_convert\_convolution} function of TVM.

% Thus, migrating tests from DL libraries to test DL compilers is a promising idea.
% \victor{why not add a figure to show the bug and the fixing commit?}
% \victor{add the test cases from Keras that can trigger this bug?}
% \qingchao{added, plz recheck}

% \jj{I think we need to discuss why the baselines cannot detect this bug. Two reviewers care about the grammar-based techniques.}\qingchao{Done, please review}
% \victor{I added some things}
It is non-trivial for those existing DL compiler testing techniques to detect this bug.
For grammar-based techniques like NNSmith~\cite{NNSmith}, developers need to manually prepare the grammar to support \texttt{Conv2DTranspose}.
As operators in DL compilers usually have complicated logic and vast space of parameters,
supporting this operator that can match the bug-triggering condition, requires extensive expert knowledge and costs.
For mutation-based techniques~\cite{tzer},
detecting this bug requires finding a seed test from a seed pool 
and applying a set of effective mutation operators,
which is also non-trivial due to large search space.
% Therefore, in this work, we conduct this study to explore an alternative and lightweight method (i.e., migration-based idea) for this challenging task.
Therefore, in this work, we explore an alternative and lightweight method (i.e., migration-based idea) for this challenging task.

%% file: 3-methodology.tex
\section{Approach}
\label{sec:methodology}
\begin{figure}
    \centering
    \includegraphics[width=.99\linewidth]{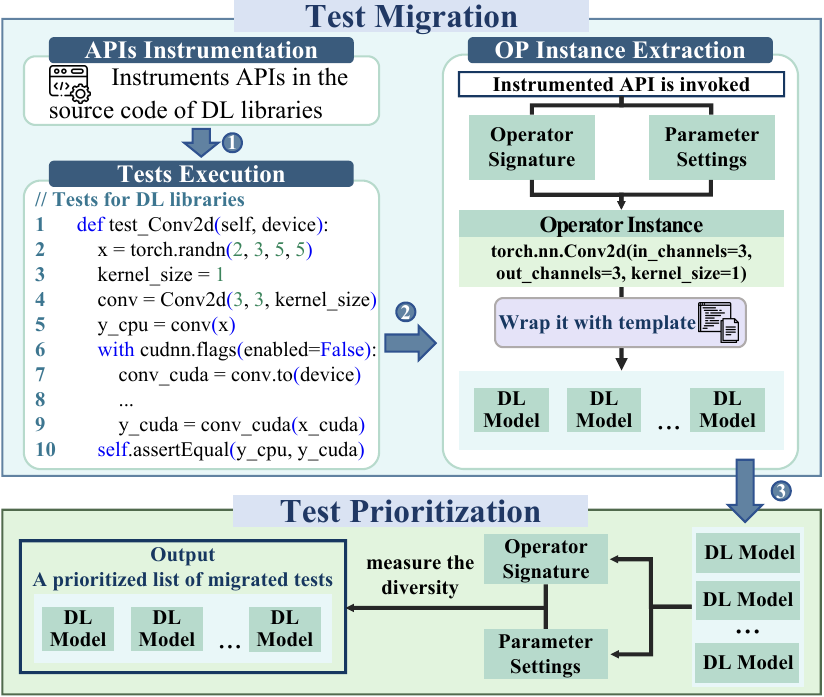}
    \caption{Workflow of \tool{}
% \jj{we don't mention migration sources.}
    }
    \label{fig:overview}
    % \vspace{-3mm}
\end{figure}

% In this work, we conducted the first empirical study to systematically investigate the feasibility and effectiveness of migrating knowledge embedded in DL library tests for testing the model loading stage.
% To support the conduct of this study, we develop a technique (i.e., \tool{}) following the migration-based idea.
% Before presenting our empirical study (the main contribution of our work), we first introduce this supporting technique.
% \jj{what's the purpose of this paragraph? there are some terms or presentations that have been discared in this version.--changed!}
% Instead of manually designing tests or developing test generators for covering operators in various DL libraries, in this work, we conduct the first exploration to enhance the testing of the model loading stage in DL compilers by migrating tests in a nearby domain, i.e., DL library testing. Such a migration-based idea can help cover operators in various DL libraries in a lightweight manner. In particular, it can relieve the burden of understanding and handling constraints in various operators when designing tests or test generators, which has been addressed well in the existing tests of DL libraries. 
With the migration-based idea, we propose a technique, called \tool{}.
The workflow of \tool{} is shown in Figure~\ref{fig:overview}, which contains two main components: test migration and test prioritization.
Specifically, \tool{} first creates tests for the model loading stage by migrating knowledge embedded in DL library tests via operator instance extraction (Section~\ref{sec:testmigration}). Due to the large number of migrated tests and the requirement of frequent test migration and execution caused by the evolution of DL libraries and compilers, \tool{} then prioritizes migrated tests based on their diversity in order to improve testing efficiency (Section~\ref{sec:tcp}). 
% That is, the tests that are more likely to detect diverse bugs can be executed earlier. 
Finally, \tool{} incorporates two test oracles to determine whether a migrated test detects a bug in the model loading stage of a DL compiler (Section~\ref{sec:oracle}).
% The workflow of \tool{} is shown in Figure~\ref{fig:overview}, which contains two main components: test migration (creating tests for the model loading stage by migrating knowledge embedded in DL library tests via operator instance extraction) and test prioritization (prioritizing migrated tests based on the diversity of operator instances).

% , to facilitate addressing the two RQs as presented in Section~\ref{sec:intro} respectively.

% \tool{} first collects and migrates the test inputs accumulated in the area of DL library testing through operator instance extraction (Section~\ref{sec:testmigration}). 
% Due to the large number of migrated test inputs and the requirement of frequent test migration and execution caused by the evolution of DL libraries and compilers, \tool{} then prioritizes migrated test inputs based on their diversity
% for investing the testing efficiency in RQ2 (Section~\ref{sec:tcp}).
% That is, the test inputs that are more likely to detect diverse bugs can be executed earlier, enabling migrated test inputs to detect more bugs within a given time budget.
% Finally, \tool{} incorporates two test oracles (will be introduced in Section~\ref{sec:rq1_set}) to determine whether a migrated test input detects a bug in the model loading stage of a DL compiler.

\subsection{Test Migration}
\label{sec:testmigration}
\subsubsection{Migration Sources}
\label{sec:data_collection}
In \tool{}, we considered both human-written tests and tool-generated tests in DL library testing as our migration sources.
Human-written tests imply expert knowledge for considering various usages of each operator under the corresponding constraints.
Tool-generated tests, on the other hand, can help explore corner cases.
In the literature, there are many DL library testing techniques~\cite{docter,deepREL,lemon}
By balancing evaluation cost and conclusion generalizability, we selected two state-of-the-art but diverse techniques (DocTer~\cite{docter} and DeepREL~\cite{deepREL}) for supporting the migration source of tool-generated tests.
DocTer extracts API constraints from official documentation and then utilizes these constraints to generate tests. 
DeepREL infers potential API relations automatically
based on API syntactic and semantic information and then synthesizes tests for invoking relational APIs.
% In total, \ul{\tool{} equipped three migration sources from DL library testing: human-written tests, DocTer-generated tests, and DeepREL-generated tests}.
In theory, \tool{} is generalizable to various DL library testing techniques and we will investigate more sources of tool-generated tests in the future.
% Earlier experiments~\cite{docter,deepREL} have shown that they can detect DL library bugs effectively.

% There are many sources of tests for DL libraries, including human-written tests and tool-generated tests, all of which can be treated as migrations sources in \tech{}.
\subsubsection{Operator Instance Extraction}
\label{sec:migration}
Although there are massive tests from the three sources, 
However, most DL library tests cannot be directly adopted to test DL compilers due to differences in their input format.
Specifically, DL compilers take DL models as inputs, while the tests for DL libraries are often in the format of Python code, most of which lack a complete model structure, as shown in Figure~\ref{fig:keras_test}. 
Hence, extracting DL models from DL library tests for DL compiler testing is non-trivial.

To achieve the goal of creating tests for the model loading stage of DL compilers by migrating knowledge embedded in DL library tests,
\tool{} uses \textit{operator instances} to bridge the migration gap.
An operator instance refers to a specific usage of an operator with a specific setting of its parameters (an example is shown in the 
% second part of 
Figure~\ref{fig:overview}).
They can be extracted from the tests for DL libraries and converted to DL models composed of a single layer for testing DL compilers.
We call such DL models \textit{single-operator models}.
Note that multiple operator instances can be extracted from one DL library test, leading to obtaining a set of single-operator models (that is, migrated tests) for DL compiler testing.

Note that the primary functionality of the model loading stage lies in converting each operator in a DL model individually into an equivalent IR, known as single-operator equivalence conversion. As a result, \tech{} naturally creates single-operator models for testing. Also, single-operator models facilitate follow-up bug de-duplication and localization.
% In our work, we also investigated the comparison with the DL models involving multiple operators from DL library testing (presented in Section~\ref{sec:result_baselines}).

\begin{figure}
    \centering
    \includegraphics[width=\linewidth]{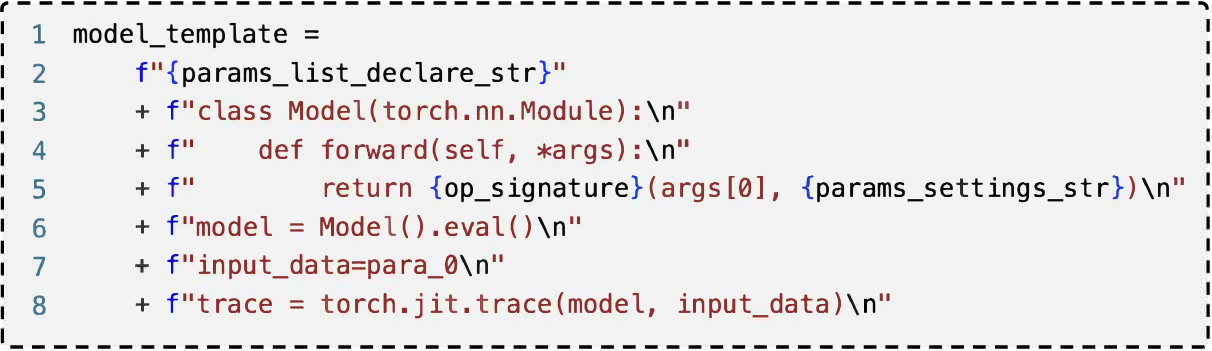}
    % \vspace{-5mm}
    \caption{Template for generating DL models under PyTorch}
    \label{fig:model_template}
    \vspace{-3mm}
\end{figure}

Specifically, \tool{} instruments APIs in the source code of DL libraries for operator instance extraction.
% \change{
% Differ from existing works~\cite{Freefuzz,docter}, only operator APIs rather than all user-level APIs in the source code of DL libraries are instrumented. The operator API is a special user-level API that can be used as a layer of a DL model. Only such operator API can be migrated to test DL compilers. We identify the operator API according to the two rules: 1) the API should take a tensor as input and give a tensor as output; 2) the API should have the \textit{trainable} attribute.}
When an instrumented API is invoked, the operator signature and its corresponding parameter values can be recorded, which collectively form an operator instance.
% The operator input is deemed one special operator parameter.
% Specifically, we hook XX operator APIs for PyTorch, XX operator APIs for Keras.
% The ONNX has composition operators which contain multiple nodes. 
% For instance, the \textit{If} operator in ONNX consists of if-node and else-nod.
% We hook the \textit{onnx.helper.make\graph} API, which includes full operator parameters settings in ONNX.
% \jj{why we need to generate inputs randomly. are we test migration, including migrating parameter settings?}
% In the test inputs for DL libraries, operator inputs are assigned random values based on the tensor shape and tensor type, such as $input=torch.randn([2, 3, 16, 130], dtype=torch.float32)$.
% Executing the tests on the instrumented code will obtain the concrete tensor values in which each element is a random value.
% To improve the efficiency of migration, we replace the concrete tensor values with tensor type and tensor shape and restore it in the format as that in the original tests. 
% \jj{I don't think this strategy is a valid migration strategy.}\qingchao{I have corrected the description, plz recheck}.
% Finally, Operator instances are obtained by wrapping the operator signature and its corresponding parameter settings.
This operator instance is then wrapped using a template as a DL model for testing the model loading stage.
% \jj{can it be used in different libraries?}\qingchao{done, plz recheck!}
Due to different DL libraries having different model construction methods, we design a model generation template for each DL library to facilitate wrapping the corresponding operator instances.
For instance, the template for PyTorch is shown in Figure~\ref{fig:model_template}, which takes an operator instance as input (Line 5) and encapsulates it in a model structure (Lines 2-5). 
An instance of the PyTorch model is then created and set to evaluation mode (Line 6). Finally, with the aid of input data, the model is serialized into deployable code (i.e., TorchScript), which can be used as the test of DL compilers.
% Based on this template, a DL model under PyTorch can be constructed, such as the example shown in Figure~\ref{fig:migrated_test_example}.
We show the used template for each DL library at our project homepage due to the space limit.
% }
% \begin{figure}
%     \centering
%     \includegraphics[width=0.95\linewidth]{figures/model_example.pdf}
%     \caption{A migrated test input under PyTorch library}
%     \label{fig:migrated_test_example}
% \end{figure}

% \jj{we should use an example to illustrate the migration process and demonstrate that a test input can produce several operator instances for DL compiler testing.}
% \qingchao{Done, plz review}The middle part of Figure~\ref{fig:overview} show the the migration process. For the tests from the migration sources, \tool{} extract the operator instances (e.g., \texttt{Threshold(input=(shape=[1,3], dtype=torch.float32),\\ threshold=0.5, value=0.9)}) by code instrumentation and wrap them using the template in Figure~\ref{fig:model_template} into a serialized DL model.
% \qingchao{can we delete this paragraph for space saving?}
% \victor{I think we have sufficient space}

\subsection{Test Prioritization}
\label{sec:tcp}
% \qingchao{R2: the approach of prioritization is verbose.}

% \victor{shall we introduce why we need to do the prioritization? I don't think we mention this in section 3? }\qingchao{added, plz take a look.}
As explained in Section~\ref{sec:intro}, the practicality of such a migration-based idea may still be hindered by significant cost consideration.
% testing overhead with the migrated-based idea can be huge, while the computational resource for testing is usually limited.
We thus incorporate test prioritization into \tool{} to detect more bugs within a given testing time budget, which facilitates investigating the efficiency improvement of this migration-based idea.
% In the literature, several test prioritization strategies have been proposed, among which coverage-based test prioritization is the most widely studied~\cite {tcp_rothermel2001prioritizing,tcp_lu2016does,tcp_li2007search}. 
% However, in the scenario of test migration from DL library testing to DL compiler testing, it is infeasible to collect test coverage on the DL compiler under test in advance. 
% Instead, collecting test coverage on the DL library is possible as the criterion of prioritizing test inputs for DL compiler testing. 
% However, the criterion is not directly aligned with the goal of improving DL compiler testing, and thus the prioritization effectiveness may be negatively affected. 
% Therefore, we do not adopt coverage-based test prioritization in \tool{} but still investigate its effectiveness in our study for sufficient evaluation (Section~\ref{sec:result_tcp}).
According to the characteristics of our scenario, we design a diversity-based test prioritization strategy in \tool{}. 
A migrated test is a single-operator model converted from an operator instance, and thus its core semantics lie in (1) the signature of the operator and (2) the setting of each parameter in the operator. 
% Intuitively, diverse semantics may indicate different testing capabilities. 
Hence, \tool{} prioritizes the set of migrated tests according to the diversity among them in terms of the two-dimensional information. 
% This is a black-box strategy and thus can be more efficient and generic than the widely-studied coverage-based strategies in general software testing.
% We also perform extensive comparisons between them in our study (Section~\ref{?}).
% Such a black-box test prioritization strategy is more efficient and generic than the widely studied coverage-based strategies.

% In the following, we first introduce how to measure the diversity in terms of operator signature in Section~\ref{sec:op_diversity}, then introduce how to measure the diversity in terms of parameter settings in Section~\ref{sec:parameter_diversity}, and finally present how to combine the two-dimensional diversity for prioritizing migrated test inputs in Section~\ref{sec:tcp_algorithem}.

\subsubsection{Diversity of Operator Signatures}
\label{sec:op_diversity}
The tests with different operator signatures mean that they have diverse semantics.
However, how to determine the order of the tests with different operator signatures is still a challenge.
In \tool{}, we address it according to the following intuitions: 
(1) the number of tests with an operator signature is large in DL library testing, indicating that this operator receives more attention potentially due to its complicated implementation logic, more corner cases in it, etc;
% The diversity of an operator signature is characterized by its occurrence frequencies in test cases of both DL libraries and DL compilers. A high frequency of an operator's signature in the tests of DL libraries but a low frequency in those of DL compilers suggests that while DL library developers place significant importance on this operator, DL compiler developers do not. It is more probable that such operator signature induces bugs in DL compilers.
% The occurrence frequency of an operator signature in DL compilers' test cases imply developers' attentio
% The more DL compiler test cases including an operator signature, indicating that this operator receives more attention potentially due to its complicated implementation logic, more corner cases in it, etc;
(2) the number of tests with an operator signature is small in the test suite equipped by the DL compiler (i.e., the test suite specific to the model loading stage), indicating that DL compiler developers still pay little attention to testing the transformation of this operator.
Hence, \tool{} assigns an operator $\textit{OP}_i$ a higher priority score 
if it occurs in the set of migrated tests for a DL library more frequently (denoted the occurrence times as $\textit{Num}\_\textit{DLL}\_\textit{OP}_i$) but occurs in the test suite equipped by the DL compiler more rarely (denoted the occurrence times as $\textit{Num}\_\textit{DLC}\_\textit{OP}_i$).
% it indicates that this operator may involve various complicated or corner cases that have been carefully considered by DL library testing, but may be overlooked by the existing DL compiler testing to some degree.
% Therefore, 
% \tool{} assigns this operator $OP_i$ a higher 
With the intuitions, the priority score of $\textit{OP}_i$ is calculated by the ratio of $\textit{Num}\_\textit{DLL}\_\textit{OP}_i$ over $\textit{Num}\_\textit{DLC}\_\textit{OP}_i$.

\subsubsection{Diversity of Parameter Settings}
\label{sec:parameter_diversity}
With the diversity of operator signatures, the tests with the same operator signature have the same priority, and thus how to further determine their priority is another challenge.
\tool{} addresses it by measuring the diversity of parameter settings for each operator.
% Specifically, an operator could have several parameters, each of which can be set to a value within its value space.
% The value space of a parameter could be large, and setting different values to a parameter may have similar testing capabilities.
Inspired by the theory of equivalence class partitioning~\cite{bhat2015equivalence}, \tool{} partitions the value space of each parameter into a set of subspaces, each of which clusters the values with potentially similar testing capabilities, by performing different considerations on different types of parameters.
% Therefore, \tool{} partitions the value space of each parameter into a set of subspaces, each of which clusters the values with potentially similar testing capabilities.
% It can be helpful to measure the diversity of parameter settings for an operator.
% The value space partitioning strategy is inspired by the idea of equivalence class partitioning~\cite{bhat2015equivalence}, which has different considerations on different types of parameters.
% Due to the limitation of pages, the detailed value space partitioning algorithm is put on our homepage.

If the value space of a parameter is \textit{a limited set of concrete values}, \tool{} treats each unique value as a unique subspace as each of them may represent a unique configuration of using this operator.
% For example, a parameter with the Boolean type only has two valid values (\ie, \texttt{true} and \texttt{false}).
% \qingchao{R1-Done: no good justification for interval partitioning of integer or float values.}
If the value space of a parameter is \textit{a range of Integers}, \tool{} pays more attention to some special values in DL~\cite{comet,powers2014boundary, dl_bug_study}, i.e., -1 (e.g., it can be used to represent the automatically-calculated dimension size in tensor shape), 0 (e.g., it can mean that no padding operation is performed when assigned to the parameter \texttt{padding}), and 1 (e.g., it can be a boundary value for the parameter \texttt{scale} in \texttt{keras.layers.Rescaling}).
% , indicated by the existing work~\cite{comet,powers2014boundary}.
Hence, \tool{} partitions them into five sets of subspaces, i.e., $(-\infty, -2], [-1], [0], [1]$, and $[2, \infty)$.
Similarly, \tool{} partitions the value space of \textit{a range of Floating number} for a parameter into three sets of subspaces, i.e., $(-\infty, 0), [0]$, and $(0, \infty)$.

If a parameter is a \textit{tensor}, which is a compound type with some typical attributes (i.e., tensor type and shape), it first partitions each tensor-type value as a unique subspace as the value space of tensor type is a limited set of concrete values.
Then, it partitions the value space of tensor shape (which belongs to the List type).
% Consequently, \tool{} partitions the value space of the Tensor according to the Tensor shape, which is in the List type.
The type of value in the List is Integer, and thus we use the Integer space partition method for it.
% \jj{is List type always in Tensor type?}\qingchao{done}
The size of the List refers to the dimension of the tensor and
% as the List type in deep learning always represents the Tensor shape.
\tool{} considers some special values~\cite{panagakis2021tensor}.
% \jj{does it indicate these specical values?}\qingchao{yes. It explains the meaning of N-order (N=2,3,4,5, >5) tensor in the CV domain.}.
% \qingchao{R1-Done: how did you determine the shape for the image and video which is according to the number of channels?}
Specifically, the dimension of the tensor with batch size ranging from zero to five can be used to represent none, scalar, vector, matrix, color image (e.g., [batch, channel, height, width]), and video (e.g., [batch, channel, depth, height, width]), respectively. 
Therefore, the value space of the dimension is divided into seven sets of subspaces, i.e., $[0], [1], [2], [3], [4], [5]$, and $[6, \infty)$.
% Since the attribute of element values is not prone to bugs, we are disregarding it.
% \victor{I cannot parse this sentence. Should it be
% ``
% For the element values not being migrated, we ignore this attribute.
% ''}\qingchao{Yes, I correct the description.}
Finally, \tool{} intersects the subspace partitioned by each attribute and thus forms the final set of subspaces for tensors.
% \begin{comment}
% \end{comment}

% value spaces of several attributes of Tensor, forming the final value partition space of Tensor.
% \jj{How to measure the diversity based on value space partition?}\qingchao{done, plz recheck}
For an operator signature, \tool{} measures the diversity score of the parameter setting for an operator instance with those of already-prioritized operator instances based on partitioned value subspaces.
Specifically, for the operator instance, \tool{} measures the percentage of parameters or pair-wise parameter combinations, whose values cover new subspaces or pair-wise subspaces over the set of already-prioritized operator instances, inspired by combinatorial testing~\cite{combination_testing}.

\subsubsection{Overall Prioritization}
\label{sec:tcp_algorithem}
% \jj{Then, how to prioritize test inputs based on the two kinds of diversity.}
With the two-dimensional diversity, \tool{} prioritizes tests based on the product of the diversity score of the operator signature and that of the parameter setting, which is called operator-instance diversity.
This calculation method may not be optimal in our scenario, and we will explore more methods in the future.
After adding the test with the maximum operator-instance diversity to the prioritized result, \tool{} updates the parameter-setting diversity of the remaining tests with the same operator signature for subsequent iterations.
The prioritization process in \tool{} is based on the Heapsort algorithm~\cite{heap_sort} due to its high efficiency ($O(\textit{nlogn})$ time complexity and $O(1)$ space complexity).

\subsection{Test Oracles}
\label{sec:oracle}
A recent study~\cite{shen2021comprehensive} revealed that most of DL compiler bugs manifested as either compiler crashes or inference inconsistencies between the original DL models and the corresponding compiled models. We thus implemented the \textbf{two test oracles} for testing the model loading stage with migrated tests.

\textit{Crash} has been widely used in compiler testing~\cite{csmith, NNSmith}, which refers to an unexpected termination of the compilation process. 
To avoid crashing a DL compiler due to invalid tests, \tool{} uses the DL library from which the tests are migrated to check their validity in advance. 
If the DL library crashes when executing a test, this test is considered invalid and thus discarded before testing the DL compiler.
In addition, the crashes that produce error messages like ``unsupported type'' and ``unsupported operator'' are disregarded as the unsupported features are often not treated as bugs.

\textit{Inference inconsistency} means that the inference results of a test (a DL model) and its corresponding compiled model via the DL compiler under test are inconsistent~\cite{ma2022hirfuzz}.
% Specifically, for a migrated test, \tool{} first executes it with the DL library corresponding to this model and obtains the inference result (i.e., a list of numbers).
% Then, \tool{} uses the DL compiler to compile the model and executes the compiled model with DL compiler runtime for obtaining the inference result.
Same as the existing work~\cite{ma2022hirfuzz,NNSmith},
we measured the inference difference between them based on Chebyshev distance~\cite{chebyshev} and used 1e-3 as the threshold to determine whether the inference results are inconsistent. 
As the DL compiler aims to achieve equivalent transformation for any DL model, if the obtained inference results from them are inconsistent, it indicates that a DL compiler bug is found.
% to reduce false positives due to floating-point issues.

%% file: 4-experiment-setup.tex
\section{EVALUATION SETUP}
\label{sec:setup}
Our evaluation aims to study two research questions:
\begin{itemize}[topsep=3pt, leftmargin=15pt]
    \item \textbf{RQ1}: To what extent can \tool{} effectively detect bugs at the model loading stage of DL compilers?
    \item \textbf{RQ2}: To what extent can the test prioritization component enable \tool{} to detect bugs earlier?
\end{itemize}

% As the first empirical study on the feasibility and effectiveness of such a migration-based idea, we mainly address two RQs (described in Section~\ref{sec:intro}) with the aid of \tech{}.

\subsection{Subjects}
% \jj{Subjects? including DL compilers and DL libraries. Explain the reason for the selection. especially why not TensorFlow?}\qingchao{done}
Following recent work~\cite{ma2022hirfuzz, tzer, NNSmith,MT_DLComp}, we performed our study on three widely-studied DL compilers, including TVM~\cite{tvm}, TensorRT~\cite{TensorRT}, and OpenVINO~\cite{OpenVINO}.
We chose the latest versions of them (i.e., TVM v0.13, TensorRT v8.6, and OpenVINO v2023.1.0), which is helpful to answer these RQs more sufficiently by detecting previously unknown bugs.
% we evaluated if migrated tests from DL libraries can detect previously unknown bugs in the latest versions of these compilers, i.e., TVM v0.13, TensorRT v8.6-release, and OpenVINO v2023.1.0-release.
% First, TVM is the most wildly-used DL compiler, which has more than 10k stars on GitHub. 
% On the other hand, TVM has the most diverse frontends, which support 13 model formats under different DL frameworks (e.g., PyTorch and ONNX).
A DL compiler usually contains multiple frontends for model loading, each of which converts DL models under a specific DL library into high-level IRs.
To allow for in-depth analysis, our evaluation focuses on the frontends of popular DL libraries, including PyTorch frontend, Keras frontend, and ONNX frontend.
Each of them handles the models under libraries that are widely used in both research and industry~\cite{dl_bug_study,comet,audee}, i.e., PyTorch, Keras, and ONNX libraries, respectively.
% and each of them represents a usage scenario of DL compilation.
% Considering TensorRT does not support Keras frontends, we only test the PyTorch and ONNX frontends for it.
% In total, eight frontends from three DL compilers are regarded as subjects in our study.
% In particular, each frontend in a DL compiler represents a unique usage scenario of the DL compiler.
% and each frontend is individual without any dependency in implementations, 
% we treated each of them  as a subject in our study.
% Due to TVM having many frontends (i.e., 13) in the model loading stage, 
% we selected the PyTorch, Keras, and ONNX frontends to test due to their popularity considering the experiment cost.
% These DL libraries have been widely used in the existing studies on DL library testing~\cite{dl_bug_study,comet,audee}.
In particular, we use Keras instead of Tensorflow as Keras is a high-level and easy-to-use interface of TensorFlow, and many DL models constructed by TensorFlow are saved in the Keras format for platform compatibility and portability at the deployment stage.
% Therefore, we used the Keras frontend instead of the TensorFlow frontend in our study, but \tool{} can be easily generalized to the frontends for other DL libraries (to be discussed in Section~\ref{sec:discuss}).
Further, as TensorRT does not support the Keras frontend, 
our evaluation covers eight frontends from three DL compilers in total.
% Noticed that \tool{} can be easily generalized to the frontends for other DL libraries (See Section~\ref{sec:discuss}).

\subsection{Baselines}
\label{sec:baselines}
We assessed the effectiveness of the migration-based idea with \tech{} by comparing it with \textbf{NNSmith}~\cite{NNSmith} and \textbf{COMET}~\cite{comet}.
Both produce DL models with multiple operators for testing, which facilitates comparison analysis with single-operator models obtained by \tech{}.

NNSmith is the state-of-the-art DL compiler testing technique, which is a grammar-based technique and can test the model loading stage as well.
Specifically, it randomly generates DL models from scratch by supporting 75 operators 
% (among 189 operators) 
of the ONNX library, which indicates that we can just study NNSmith on the ONNX frontend.
% As the test migration aims to enhance the testing of the model loading stage of the DL compiler, we firstly compared it with the state-of-the-art DL compiler testing technique (i.e., \textbf{NNSmith}~\cite{NNSmith}), which is a grammar-based technique and can test the model loading stage as well.
% The remaining DL compiler testing techniques (i.e., HirGen, Tzer, TVMFuzz) skip this stage by constructing high-level IRs or low-level IRs as test inputs.
% \subsubsection{Test approach for DL compilers}
% \textit{NNSmith} is the state-of-the-art test method for DL compilers. 
% Specifically, NNSmith randomly generates DL models by supporting 75 operators (among 189 operators) of the ONNX library, which indicates that we can just compare \tool{} with NNSmith on the ONNX frontend.
Even though most of DL library tests cannot be directly adopted to test the model loading stage, we have to use \tech{} to support the migration.
There are also some DL library testing techniques that can directly generate DL models to satisfy the input format of DL compilers.
We also studied such a state-of-the-art technique, i.e., COMET, which designs a set of mutation operators and a coverage-based search-based algorithm to generate diverse models for testing DL libraries.

In RQ2, we investigated the efficiency improvement of such the migration-based idea by evaluating the effectiveness of the test prioritization strategy in \tech{}.
Here, we considered some test prioritization strategies commonly used in general software testing for comparisons.

\begin{itemize}[topsep=0pt, leftmargin=*]
    \item \textbf{Random}. 
    The migrated tests are  randomly ordered, serving
    % can be regarded 
    as the baseline without special test prioritization.
    \item \textbf{FAST}~\cite{fast}, which treats each test as a string, and adopts the data mining algorithms (i.e., minhashing and locality-sensitive hashing algorithms~\cite{jafari2021survey}) to accelerate the process of finding diverse tests by converting each string to a k-shingle (the set of its substrings of length k). 

    \item \textbf{Total}-coverage-based prioritization, which prioritizes migrated tests based on the number of program elements (in the frontend under test) covered by each test. 
    Here, we used statements as the representative program elements following existing work~\cite{tcp_rothermel2001prioritizing,tcp_zhou2021parallel,tcp_chen2023exploring,tcp_chen2018optimizing}.
    % , and collected DL library coverage (rather than DL compiler coverage) due to our test migration scenario as explained in Section~\ref{sec:tcp}.
    \item \textbf{Additional}-coverage-based prioritization, which prioritizes migrated tests based on the number of covered statements that are not covered by the existing prioritized ones. 
    % Similarly, we collected statement coverage in the frontend under test as the prioritization criterion.
\end{itemize}

% For ease of presentation, we call the four variants of \tool{} with different test prioritization strategies \textbf{\tool$_{\textit{random}}$}, \textbf{\tool$_{\textit{FAST}}$}, \textbf{\tool$_{\textit{total}}$}, and \textbf{\tool$_{\textit{additional}}$}, respectively.
FAST is the state-of-the-art black-box strategy, while coverage-based prioritization is the most widely studied white-box strategy.
The prioritization strategy in \tool{} and the random strategy are black-box.
For coverage-based strategies, Coverage.py~\cite{coverage_tool} is used to collect statement coverage in frontends.
For FAST, we re-used its released implementation~\cite{fast}.

\subsection{Metrics}
\label{sec:metrics}

We counted \textbf{the number of detected bugs} 
% within a given testing time budget 
as the metric of evaluating the test effectiveness.
% A recent study~\cite{shen2021comprehensive} points out that most of historical DL compiler bugs manifested as either compiler crashes or inference inconsistencies between the original DL models and the corresponding compiled models.
% Leveraging this finding, we design the following two test oracles to determine whether a test input uncovers a bug:
% 
% 
% Note that randomness may exist in a test input, potentially resulting in false positives, and thus \tool{} executes each test input twice with the DL library.
% If the inference results from the two executions are inconsistent, we directly discard it before DL compiler testing.
% \ul{De-duplication}:
During the testing process,
it is possible that
some of the test failures
are triggered by the same root cause.
Hence, it is important to \ul{de-duplicate} them and count the number of unique bugs.

In DL compiler, each operator in the model loading stage comprises a conversion function that is responsible for converting it into the equivalent high-level IR.
As the migrated test is a single-operator model, it is convenient to determine the conversion function responsible to the operator in a failure-triggering test.
Therefore, based on the identified conversion function for each test failure, we de-duplicated the test failures to obtain unique bugs.
Here, we did not use the operator in each failure-triggering test for de-duplication, as different operators may be handled by the same conversion function in the DL compilers.
For example, {\tt AveragePooling2D} and {\tt MaxPooling2D} are two different operators in the Keras library, but they are converted by the same function (i.e., {\tt \_convert\_pooling}) in TVM.
% Hence, the de-duplication method based on the identified transformation functions can be more accurate.
As the tests generated by NNSmith and COMET are DL models with multiple operators, we manually de-duplicated their test failures 
% with the aid of crash messages 
following the original papers~\cite{NNSmith,comet}.

All bugs are detected on the latest versions of DL compilers, and thus we created a bug report for each unique bug and then submitted it to project maintainers.
We counted the number of confirmed or fixed bugs by developers.
Based on the feedback from developers, 
% our de-duplication methods do not produce any false positives.
all of our submitted bugs that have been confirmed are unique. This indicates the accuracy of our de-duplication method. %, confirmed by project maintainers, indicating the accuracy of our de-duplication methods to some degree.

Besides, \tool{} includes a test prioritization component to improve the testing efficiency, and thus it is also important to investigate the testing efficiency of each technique. 
Here, we used two metrics to evaluate the effectiveness 
of each prioritization strategy.
First, we measured \textbf{the time spent on detecting each bug}.
The shorter the time is, the more effective the strategy is.
Second, We adopted the widely-used metric of evaluating test prioritization, i.e., \textbf{APFD} (Average Percentage of Faults Detected)~\cite{apfd}, to compare various test prioritization strategies following many existing 
studies~\cite{tcp_henard2016comparing,tcp_chen2023exploring,chen2016test,lou2019survey}.
The calculation of APFD is shown in Formula~\ref{eq:apfd}:
\begin{equation}
\label{eq:apfd}
    APFD = 1 - \frac{\sum_{i=1}^{m}(p_i)}{n \cdot m} + \frac{1}{2n}    
\end{equation}
% APFD is calculated as $\textit{APFD} = 1 - \frac{\sum_{i=1}^{m}(p_i)}{m \cdot n} + \frac{1}{2n}$,
where $m$ is the total number of detected bugs, $n$ is the total number of tests, $p_i$ is the rank of the first test in the prioritized result that detects the $i^{\textit{th}}$ bugs.
A larger APFD value indicates a more effective strategy.

\subsection{Implementations}
\label{sec:implement}
We collected the test suite equipped by PyTorch v1.7, Keras v2.3, and ONNX 1.8 as the migration source of human-written tests for the corresponding frontends of the three DL compilers, respectively.
We collected 32,378, 20,992, and 1,014 tests from the three test suites, respectively.
% DocTer and DeepREL provide the migration sources of tool-generated test inputs in \tool{}.
We used the implementations of DocTer and DeepREL released by their works~\cite{docter,deepREL}.
As neither of them supports test generation for the ONNX library, we exclude them when using \tool{} to test the ONNX frontend in DL compilers.
For the PyTorch and Keras frontends, we used DocTer and DeepREL to generate the same number of tests as the corresponding human-written tests respectively, which can help compare the three migration sources fairly.
All experiments were conducted on an Ubuntu 18.04 server with Intel Xeon CPU, NVIDIA GTX1080Ti GPU, and 128G RAM.

% Weight quantization~\cite{hou2018loss} is commonly used by DL compilers to reduce model sizes by converting weights from a given type to {\tt {float16}} or even {\tt int8} type.
% However, weight quantization inevitably leads to a loss in accuracy~\cite{dflare}, which is usually not considered as a bug.
% To mitigate the inconsistencies resulting from weight quantization, we switched off the quantization in DL compilers.

\subsection{Process}
For each studied frontend, we obtained a set of migrated sets from the three migration sources with the aid of \tech{}.
We then tested each frontend with these migrated tests and recorded whether a test triggered a failure or not and the time spent on each test.
% \qingchao{\textbf{clarify the generation time}}
For fair comparison, we applied NNSmith and COMET to generate DL models for testing each frontend for the same time budget as that used by \tool{} (including the time spent on test generation, migration, prioritization, and execution by \tool{}), respectively.

To answer RQ2, for each frontend, we constructed four variants of \tool{} by replacing its diversity-based test prioritization strategy with each of the four compared strategies, respectively.
By applying each variant to test each frontend, we recorded the test result and testing time of each test, and calculated the APFD value.
To reduce the influence of randomness and the running environment, we repeated our experiments for five times and calculated average results.

%% file: 5-evaluation.tex
\section{Results and Analysis}
\label{sec:evaluation}

% In this section, we present and analyze the results to answer the RQs we designed in Section~\ref{sec:setup}.

\subsection{RQ1: Effectiveness}
\label{sec:result_bugs}
% For each studied frontend, we first applied \tool{} to migrate test inputs from the three migration sources.
% For fair comparison in RQ1, we then applied NNSmith and COMET to test the ONNX frontend of the DL compilers for the same testing time as \tool{} (including the time spent on migration and execution by \tool{}) on testing the ONNX frontend separately.

% \jj{1. show the number of bugs detected by \tool{}, and some details on detected bugs, we can also show some interesting cases for detailed illustration;
% 2. show the comparison results with NNSmith;
% 3. analyze the results of \tool{} from different sources;
% 4. false positives.}\qingchao{done, plz check!}

% \begin{table}[]
% \caption{Number of bugs detected by \tool{}}
% \vspace{-1mm}
% \label{tab:bugs_detail}
% \begin{tabular}{ccccc}
% \toprule
% \textbf{Status} & \textbf{PyTorch} & \textbf{Keras} & \textbf{ONNX} & \textbf{Total} \\ \midrule
% Fixed           & 10               & 18             & 2             & 30             \\
% Confirmed       & 0                & 7              & 5             & 12             \\
% Awaiting        & 21               & 8              & 7             & 36             \\ \midrule
% Total           & 31               & 33             & 14            & 78             \\ \bottomrule
% \end{tabular}
% \end{table}
\begin{table}[]
\caption{Number of bugs detected by \tool{}. ``--'' means not applicable since TensorRT does not support Keras models.}
% \vspace{-3mm}
\label{tab:bugs_detail}
 \renewcommand{\arraystretch}{0.7}
\resizebox{0.95\linewidth}{!}{

\begin{tabular}{cccccc}
\toprule
\textbf{Frontend} & \textbf{Status} & \textbf{TVM} & \textbf{TensorRT} & \multicolumn{1}{l}{\textbf{OpenVINO}} & \textbf{Total} \\ \midrule
\multirow{3}{*}{PyTorch} & Fixed     & 9 & 0  & 7  & 16  \\
                         & Confirmed & 0  & 6  & 10 & 16   \\
                         & Awaiting  & 21 & 25 & 7  & 53  \\ \midrule
\multirow{3}{*}{Keras}   & Fixed     & 20 & --  & 6  & 26  \\
                         & Confirmed & 5  & --  & 2  & 7  \\
                         & Awaiting  & 10  & --  & 3  & 13  \\ \midrule
\multirow{3}{*}{ONNX}    & Fixed     & 2  & 4  & 2  & 8   \\
                         & Confirmed & 5  & 5  & 7  & 17  \\
                         & Awaiting  & 7  & 3  & 4  & 14  \\ \midrule
\multicolumn{2}{c}{Total}            & 79 & 43 & 48 & 170 \\ \bottomrule
\end{tabular}}
% \vspace{-4mm}
\end{table}

% \begin{table}[]
% \caption{DL Operators Coverage by \tool{}}
% \label{tab:op_cov}
% \begin{tabular}{cccc}
% \toprule
%               & PyTorch & Keras   & ONNX    \\ \midrule
% Covered OPs   & 189     & 99      & 177     \\
% Total OPs     & 267     & 113     & 189     \\ \midrule
% Coverage Rate & 70.79\% & 87.61\% & 93.65\% \\ \bottomrule
% \end{tabular}
% \end{table}
\subsubsection{Bug Detection}

Table~\ref{tab:bugs_detail} shows the number of bugs detected by tests migrated from the testing of PyTorch, Keras, ONNX libraries with the aid of \tool{}, respectively.
% They take different frontends (i.e., PyTorch, Keras, ONNX frontends) as the testing entries for TVM.
% In total, \tool{} detects \TotalBugsNum{} previously unknown bugs in TVM, including \PTBugsNum{} on the Pytorch frontend, \KerasBugsNum{} on the Keras frontend, and \ONNXBugsNum{} on the ONNX frontend.
In total, \TotalBugsNum{} previously unknown bugs are detected, including \TVMBugsNum{}, \TRTBugsNum{}, and \OVBugsNum{} on TVM, TensorRT, and OpenVINO, respectively.
% \tool{} take different frontends (i.e., PyTorch, Keras, ONNX frontends) as the testing entries for each DL compiler and uncover \PTBugsNum{} bugs on the Pytorch frontend, \KerasBugsNum{} bugs on the Keras frontend, and \ONNXBugsNum{} bugs on the ONNX frontend.
\ConfirmedBugsNum{} bugs of them have been confirmed or fixed by developers, and the remaining bugs are being investigated by developers.
The ratio of confirmed or fixed bugs is high for most of the subjects,
except the PyTorch frontend for TensorRT, since its developers are inactive.
% , resulting in a low ratio of confirmed or fixed bugs.

% The number of confirmed or fixed bugs is high in most frontends except the PyTorch frontend of TensorRT and the ONNX frontend of OpenVINO whose person in charge has a low level of activity in the project.

% Since each migrated test input by \tool{} is a single-operator model, which is hard to activate various optimizations due to its simple model structure, almost all the bugs detected by \tool{} are DL compiler frontend bugs that have been confirmed by our manual analysis.
% \victor{what is the purpose of the above sentence? I got confused.}
% \victor{we want to say it is simple? or do we want to say they are confirmed by us? I did not get the relation behind ``since''}
% Moreover, after the investigation by developers, all the \ConfirmedBugsNum{} bugs that have been confirmed or fixed are indeed frontend bugs.
After investigations by developers, apart from three optimization bugs, the remaining \ConfirmedFrontBugsNum{} bugs that have been confirmed or fixed are frontend bugs.
This is aligned with the goal of enhancing the testing of the model loading stage via test migration.
Specifically, 
each migrated test by \tool{} is a single-operator model, which can trigger various logic in the model loading stage,
but the bugs in the optimization stages
often involve more complicated models~\cite{ma2022hirfuzz}.

% \find{The tests for DL libraries can be effectively migrated with the aid of \tech{} to test the model loading stage of DL compilers.
% \TotalBugsNum{} previously unknown bugs are detected in the three DL compilers, complementing existing DL compiler tests well.
% }

\subsubsection{Root Causes of Detected Bugs}
According to the feedback from developers on the \ConfirmedBugsNum{} confirmed or fixed bugs and their patches, these bugs are caused by diverse root causes, which cover all root cause categories summarized on historical bugs in the model loading stage~\cite{shen2021comprehensive}.
Among the \ConfirmedBugsNum{} bugs, 28 bugs are caused by Tensor Shape Problem, 18 bugs are due to Type Problem, 17 bugs are Incorrect Code Logic, 13 bugs are due to Incorrect Exception Handling, 5 bugs are due to Incompatibility, 4 bugs are due to Incorrect Assignment, 3 bugs are Incorrect Numerical Computation, 1 bug is due to Concurrency, and 1 bug is due to Typo.
These root causes are classified based on the existing study on DL compiler bug~\cite{shen2021comprehensive}.
% Due to the diversity of our detected bugs, our 
% Such diverse bugs detected by \tool{}
% are an effective supplement to the existing bug study about DL compilers~\cite{shen2021comprehensive} which only consists of 116 bugs in the model loading stage.
% \victor{usually we don't use the word ``good''
% in academic writing}

\begin{figure}[]
    \centering
    \includegraphics[width=.95\linewidth]{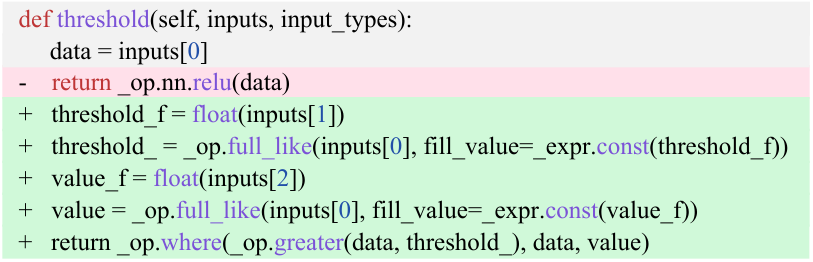}
    % \vspace{-2mm}
    \caption{Patch for an Incorrect Code Logic bug}
    \label{fig:bug_inconsistency}
    % \vspace{-1mm}
\end{figure}

Next, we present two bugs detected by the migrated tests.
% \jj{we need to explain the root cause of them, how to trigger them, and how to fix them. Please sample some interesting or critical bugs.}
% \qingchao{Done, plz check}
% \jj{to change an example or remove an example.}\qingchao{done, add a new example shown below!}
Figure~\ref{fig:bug_inconsistency} depicts an Incorrect Code Logic bug in the PyTorch frontend of TVM~\cite{bug_logic}.
% \footnote{https://github.com/apache/tvm/issues/14805}.
In this bug, the \texttt{Threshold} operator from PyTorch was converted into the high-level IR of \texttt{\_op.nn.relu}, which differs in its computation logic.
Any non-zero value assigned to the parameter \texttt{threshold} or \texttt{value} of the \texttt{Threshold} operator will lead to wrong inference results. 
% \victor{it seems the threshold only has parameter inputs and input types? 
% where is the threshold and value?}
% We discovered dozens of DL models on GitHub meet the bug-triggering conditions, indicating this bug is hazardous.
Indeed, a migrated test containing the operator instance \texttt{torch.nn.Threshold(threshold=2, value=1)}, helps trigger this bug during compilation. 
A patch~\cite{patch_logic}
% \footnote{https://github.com/apache/tvm/pull/14820} 
was committed to fix it by correcting the conversion logic for the \texttt{Threshold} operator as shown in Figure~\ref{fig:bug_inconsistency}.
\begin{figure}
    \centering
    \vspace{-1mm}\includegraphics[width=.95\linewidth]{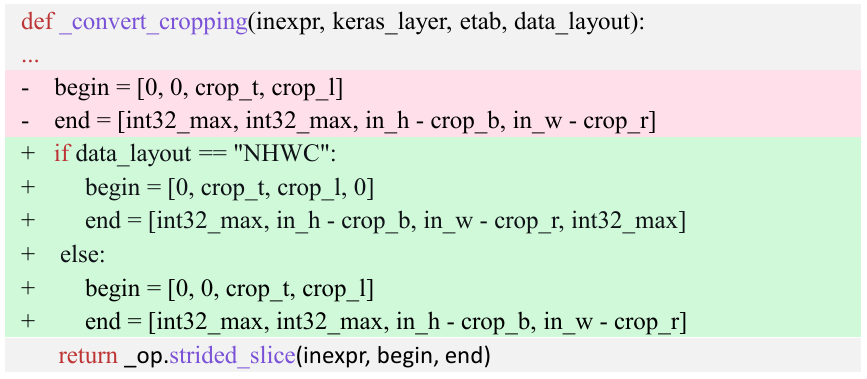}
    % \vspace{-2.5mm}
    \caption{Patch for a Tensor Shape bug}
    \vspace{-3mm}
    \label{fig:bug_shape}
\end{figure}

Figure~\ref{fig:bug_shape} shows another bug~\cite{patch_shape}
% \footnote{https://github.com/apache/tvm/pull/15053} 
caused by Tensor Shape Problem in the Keras frontend.
When converting the \texttt{Cropping2D} operator, TVM always considers the data layout to be \texttt{NCHW} (e.g.,  channel first), but \texttt{NHWC} (e.g., channel last) is also a common data layout.
When TVM loads the model containing the \texttt{Cropping2D} operator and sets the parameter \textit{data\_format} to {\tt channels\_last}, which means the data layout is in the \texttt{NHWC} format, this bug can be triggered and lead to wrong inference results.
This bug has been fixed using different calculation logic for different layouts.

\subsubsection{Comparison with NNSmith and COMET}
\label{sec:result_baselines}
% To further evaluate \tool{} is an effective method, we compare it with the state-of-the-art NNSmith for DL compiler testing. The experiment process is described in Section~\ref{sec:process}.

% As NNSmith just generates ONNX models, we compared \tool{}, NNSmith, and COMET by taking the ONNX frontend as the representative.
During the same testing time, the migrated tests by \tech{} detect \TVMBugsNum{}, \TRTBugsNum{}, \OVBugsNum{} bugs, while NNSmith detects \NNSmithTVMBugsNum{}, \NNSmithTRTBugsNum{}, \NNSmithOVBugsNum{} bugs and COMET detects \COMETTVMBugsNum{}, \COMETTRTBugsNum{}, \COMETOVBugsNum{} bugs in TVM, TensorRT, OpenVINO, respectively.
% The majority of detected bugs by NNSmith and COMET are in optimization stages (i.e., 15 of 18 for NNSmith and 7 of 10 for COMET), due to constructing more complex models involving multiple operators.
% All frontend bugs detected by them are also detected by the migrated tests by \tech{}.
15 of 18 bugs detected by NNSmith and 7 of 10 bugs detected by COMET are unique, which were not detected by \tech{}. All these unique bugs are in the optimization stages. 
Besides, all frontend bugs detected by NNSmish and COMET are also detected by \tech{}.
\tech{} detected 164 unique bugs that were not detected by NNSmith and COMET, 156 of which are frontend bugs.
The results demonstrate the superiority of \tech{} in testing the model loading stage and the complementarity between \tech{} and the existing DL compiler testing techniques (NNSmith and COMET).
The major reason for the superiority of migrated tests by \tech{} over NNSmith and COMET in testing the model loading stage is that the latter two support only 75 and 72 operators while the former covers 477 operators in a lightweight manner. 
This implies a correlation between operator coverage and bug detection, \ie, higher operator coverage in testing
is likely to detect more bugs in the model loading stage.

\subsubsection{Contribution of Different Migration Sources and Different Test Oracles}
\label{sec:result_sources}

\begin{figure}[t]
    \centering
    \includegraphics[width=0.98\linewidth]{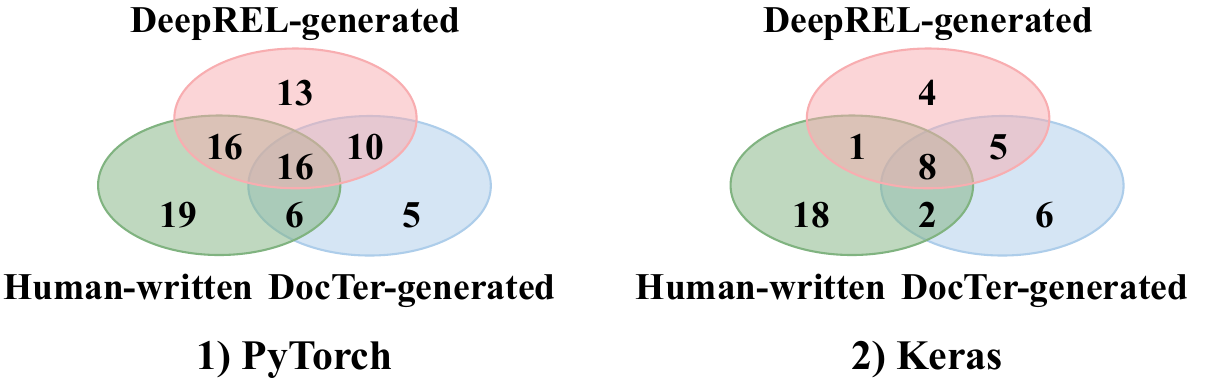}
    % \vspace{-2mm}
    \caption{Bug detection comparison among different sources}
    % \victor{use a smaller circle to save some space}}
    % \vspace{-3mm}
    \label{fig:vn_graph_bugs}
\end{figure}

In this work, \tool{} considers three migration sources (i.e., tests documented in DL libraries, and the tests generated by two recent fuzzers)
and designs two test oracles (i.e., crash and inference inconsistency).
Here, we analyzed the contribution of each migration source
as well as each test oracle.
Figure~\ref{fig:vn_graph_bugs} shows the bug detection results for each migration source in \tool{}, which do not include the results on the ONNX frontend as DocTer and DeepREL do not support the testing of the ONNX library.
% and thus only one migration source is considered by \tool{} for the ONNX frontend.
From Figure~\ref{fig:vn_graph_bugs}, all three migration sources contribute to detecting a certain number of unique bugs in PyTorch and Keras frontends, showing the complementarity among them in testing the model loading stage.
Among three sources,
the migration source of human-written tests detects the most bugs in both PyTorch and Keras frontends.
% This is because the human-written tests cover more than twice the number of operators than the other two.
Specifically, the migrated tests from human-written tests cover 169 PyTorch operators and 131 Keras operators, while the migrated tests from DocTer cover 65 PyTorch operators and 53 Keras operators, and the migrated tests from DeepREL cover 59 PyTorch operators and 26 Keras operators.

% \find{Different migration sources show the complementarity in DL compiler testing. The migration source of human-written tests is more effective than the other two sources of tool-generated tests in bug detection due to covering more operators.}

% \subsubsection{Contribution of Different Test Oracles}
Through further analysis, we found that among the \TotalBugsNum{} bugs detected by \tool{}, \TotalCrashBugs{} are detected by the test oracle of crash (including \TotalConfirmedCrashBugs{} confirmed/fixed bugs) while \TotalWrongBugs{} are detected by the test oracle of inference inconsistency (including \TotalConfirmedWrongBugs{} confirmed/fixed bugs).
The results demonstrate that the two test oracles are complementary for detecting DL compiler bugs with \tool{}.

\subsubsection{False Positives}
\label{sec:rq1_fp}
There are only 9 false positives produced by the migrated tests in total.
All of them are caused by the test oracle of inference inconsistency between the DL compiler and the corresponding DL library.
Specifically, two of them are the bugs in the Keras library rather than the DL compiler, which have been fixed in the latest version of Keras.
% They were identified by our manual analysis, and thus we did not submit them to TVM developers.
In this work, we assume that the bug occurs at the DL compiler when there is an inference inconsistency between a DL library and a DL compiler, thus leading to the two false positives.
% \qingchao{R3-Done:why the fp rate is acceptable in practice?}
% The remaining 

Three false positives are due to the undefined behaviors in the \texttt{Mod}, \texttt{RoiAlign}, \texttt{Trilu} operators.
For example, \textit{Mod} takes the dividend tensor and the divisor tensor as inputs and produces the remainder of them.
If the dividend is zero, the result will be platform-dependent.
That is, this false positive is caused by the undefined behavior at division by zero, which is also meaningful as the OpenVINO developer commented: ``It is a good catch. We will count on this issue in case we face undefined behavior later.''
% Although \tool{} catches a few false positives, each of them is valuable.

The remaining 4 false positives are due to randomness in operators (e.g., \texttt{Bernoulli} and \texttt{RandomUniformLike}).
For example, \texttt{Bernoulli} takes as input a tensor containing probabilities and draws the binary random number from a Bernoulli distribution.
\texttt{RandomUniformLike} generates a tensor with random values drawn from a uniform distribution.
The false positives caused by randomness may be filtered out by checking whether these inference inconsistencies also exist between different versions of the DL library.

% For the false positives, bugs in DL libraries and undefined behavior are useful while randomness is harmful.
% Therefore, setting a strategy to fill out the harmful false positives automatically is necessary.
% For example,  the test input should be executed multiple under the same conditions to judge if the result of the test input is deterministic.
% Indeterminate test inputs should avoid using the test oracle of \textit{inference inconsistency}.

% \find{Testing with the migrated tests by \tech{} produces few false positives, which are all caused by the test oracle of inference inconsistency arising from diverse reasons (including undefined behaviors and randomness). 
% }

\subsection{RQ2: Efficiency}
\label{sec:result_tcp}

\begin{figure*}
    % \vspace{-4mm}
    \centering
    \includegraphics[width=0.98\linewidth]{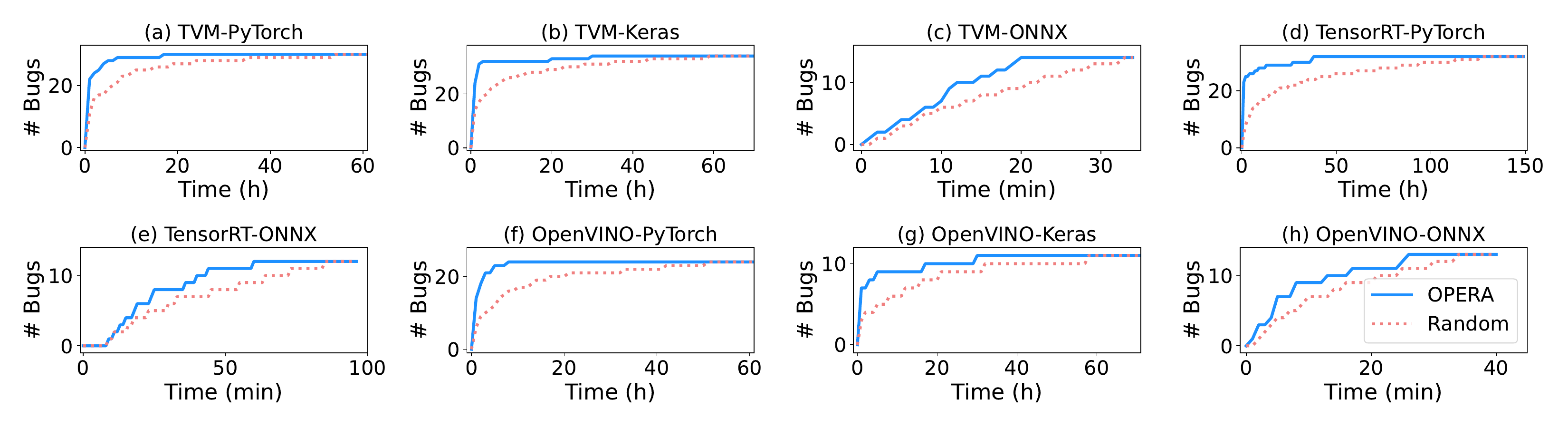}
    % \vspace{-2mm}
    \caption{Trend of bug detection effectiveness with the testing process proceeding}
    \label{fig:tcp_trends}
    % \vspace{-3mm}
\end{figure*}

We explored the efficiency improvement of our migration-based idea by investigating whether the test prioritization component in \tech{} can help detect more bugs with a given testing time budget.
% Here, we compared \tool{} with its variant without prioritization (\tool$_{\textit{pure}}$).
Figure~\ref{fig:tcp_trends} shows the number of detected bugs by \tool{} and its variant without special prioritization (\tool$_{\textit{random}}$) with the testing process proceeding on each subject.
As \tool{} has extra time spent on test prioritization but \tool$_{\textit{random}}$ does not, we included its prioritization time into the testing time of \tool{} for fair comparison.
Note that the total time cost across different subjects is inconsistent due to the varying number of migrated tests from different DL libraries (presented in Section~\ref{sec:implement}) and the differing compilation time across DL compilers.

From Figure~\ref{fig:tcp_trends}, \tool{} always detects more bugs than \tool$_{\textit{random}}$ regardless of the given testing time budget.
% except the first two minutes due to the extra prioritization time of \tool{}
% (achieving similar bug-detection effectiveness during the initial testing time).
% \jj{why onnx results are bad in the initial testing time.} 
% \st{except the initial testing time due to the extra prioritization time of \tool{}}
% \qingchao{even in the initial testing time due to the negligible extra prioritization time of \tool{}.
% The extra prioritization time cost for all migrated test inputs of PyTorch, Keras, ONNX frontend is 86.06, 18.22, 10.64 seconds, respectively.
% The random method has a good effects in the initial 3 minutes testing time for ONNX due to ...}
% That is, \tool{} is more effective regardless of the given testing time budget.
% In particular, \tool{} spends 12.72, 17.81, 0.26 hours on detecting all the targeted bugs in PyTorch, Keras, ONNX frontends while \tool$_{\textit{random}}$ spends 34.35, 43.37, 0.53 hours respectively.
In particular, \tool{} spends 16.21, 29.31, 0.32, 37.33, 1.00, 7.95, 29.81, 0.42 hours on detecting all the bugs found in the experiment presented in Section~\ref{sec:result_bugs} on each subject (in the order shown in Figure~\ref{fig:tcp_trends}),
while \tool$_{\textit{random}}$ spends 53.01, 58.56, 0.54, 125.46, 1.40, 50.76, 57.22, 0.56 hours respectively.
On average, the application of test prioritization in \tool{} leads to a more than 55.88\% reduction in time spent on bug detection,
% In total, the application of test prioritization in \tool{} leads to a more than 76\% reduction in time spent on bug detection.
confirming the contribution of the test prioritization component of \tech{} in efficiency improvement.
% \qingchao{The total time reduction rate (e.g., 76\%)is higher. could we show the total time reduction rate instead of the average reduction rate?}
% \jj{why \tool{} performs worse in the beginning on TensorRT-ONNX?}
% \qingchao{Solved, the bad result due to the some equipped test case not be collected, we have solved it }
% \qingchao{todo: show the figure}
% Figure~\ref{fig:tcp_trends} depicts the time spent on detecting each bug among \tool{}, $\tool{}_{random}$, and NNSmith\qingchao{will be added.}.
% Compared with NNSmith, \tool{} and  $\tool{}_{random}$ have an extra cost of \MigrateTime{} hours on migration and \TCPTime{} minutes on prioritization test inputs.
% When the test migration and prioritization are finished, \tool{} can uncover many unique bugs quickly.
% \tool{} can detect all the bugs within the time XX, XX, and XX hours for the 3 DL \tool{} saved XX\% test execution time for detecting all bugs from all migrated test inputs.

\begin{table}[]
\caption{Comparison among different prioritization strategies in terms of APFD}
% \vspace{-3mm}
\label{tab:APFD}
\resizebox{\linewidth}{!}{%
\renewcommand{\arraystretch}{1.1}
\begin{tabular}[width=\linewidth]{ccccccc}
\toprule
\textbf{Compiler} & \textbf{Frontend} & \textbf{\tool{}} & \textbf{Random} & \textbf{FAST} & \textbf{Total} & \textbf{Additional} \\ \midrule
\multirow{3}{*}{TVM}       & PyTorch & \textbf{0.984}         & 0.913                  & 0.930 & 0.815 & 0.778 \\
                           & Keras   & \textbf{0.976}         & 0.905                  & 0.906 & 0.691 & 0.868 \\
                           & ONNX    & \textbf{0.727}         & 0.577                  & 0.487 & 0.316 & 0.575 \\ \midrule
\multirow{3}{*}{TensorRT}  & PyTorch & \textbf{0.982}         & 0.867                  & 0.930 & 0.706 & 0.854 \\
                           & Keras   &      --                &      --                & --    & --    & --    \\
                           & ONNX    & \textbf{0.768}         & 0.645                  & 0.628 & 0.663 & 0.334 \\ \midrule
\multirow{3}{*}{OpenVINO}  & PyTorch & \textbf{0.983}         & 0.877                  & 0.927 & 0.608 & 0.670 \\
                           & Keras   & \textbf{0.946}         & 0.850                  & 0.919 & 0.750 & 0.740 \\
                           & ONNX    & \textbf{0.816}         & 0.716                  & 0.689 & 0.323 & 0.416 \\ \midrule
\multicolumn{2}{c}{\textit{Average}} & \textbf{0.898}         & 0.794                  & 0.802 & 0.609 & 0.654 \\ \hline
\end{tabular}
}
% \vspace{-5mm}
\end{table}

% \find{The test efficiency of the migrated tests can be largely improved via the test prioritization component in \tech{}, achieving an average of 55.88\% reduction in time spent on bug detection.}

We also compared the test prioritization strategy designed in \tool{} and several existing test prioritization strategies in general software testing
, based on all the migrated tests and all the detected bugs in the experiment presented in Section~\ref{sec:result_bugs}.
% \qingchao{Done: explain the conditions under which APFD is being evaluated.}
% In this experiment, we used the migrated tests implemented in section~\ref{sec:implement} and all the detected bugs in Table~\ref{tab:bugs_detail} for evaluation.
Table~\ref{tab:APFD} shows the comparison results among \tool{} and its variants with four existing test prioritization strategies (introduced in Section~\ref{sec:baselines}) in terms of APFD.
% Due to the space limit, we put the detailed results (like Figure~\ref{fig:tcp_trends}) at our project homepage
We found that \tool{} performs the best among all the test prioritization strategies on all eight subjects.
On average across all eight subjects, the APFD value of \tool{} is \AvgAPFD{} with the improvement of \APFDRandomImprove{}, \APFDFastImprove{}, \APFDTotalImprove{}, \APFDAdditionImprove{} over \tech{} with random order, FAST, total-coverage-based test prioritization, and additional-coverage-based test prioritization, respectively.
The results demonstrate the effectiveness of the test prioritization strategy designed in \tool{}.
This also indicates that designing a test prioritization strategy specific to this migration scenario is more effective than general strategies regardless of white-box or black-box strategies.

The test prioritization strategy in \tech{} contains two aspects: diversity of operator signature and diversity of parameter settings. We also performed an \textit{ablation study} to measure the contribution of each aspect by constructing two variants of \tech{}: \tool$_{\textit{op}}$ (only using diversity of operator signature to prioritize tests) and \tool$_{\textit{para}}$ (only using diversity of parameter settings to prioritize tests).
On average across all eight subjects, the APFD values of \tool$_{\textit{op}}$ and \tool$_{\textit{para}}$ are 0.613 and 0.832, while the APFD value of \tech{} is \AvgAPFD{}, demonstrating the contribution of each aspect.

%% file: 5.2-discussion.tex
\section{Discussion}
\label{sec:discuss}

% \subsection{Implications}
% \label{implication}
% We discuss the implications on the detection of DL compiler bugs according to our findings and the design of test oracle according to false positives.

% \smallskip
% \noindent\textbf{Based on Findings}
% \myparagraph{Based on Findings}
% From January 2017 to March 2024, more than 13k code changes were committed in TVM.`
% Regression testing is a vital task for DL compilers due to their rapid development.
\noindent\textbf{Generalizability.}
We evaluated \tech{} by migrating testing knowledge from three DL libraries to test eight frontends of three DL compilers.
The consistent conclusions demonstrate the generalizability of it.
Hence, it is promising to leverage \tech{} for testing more frontends of more DL compilers in the future.
Besides, the rich set of migrated tests can be directly used to test the other software taking DL models as inputs (e.g., model converters like MMdnn~\cite{mmdnn} that converts a DL model under one DL library into the equivalent model under another DL library) without any adaptation.
The richness and diversity of these migrated tests may be also helpful for them.

% According to Findings 1 and 3, the tests for DL libraries can be migrated to test the model loading stage of DL compilers effectively.
% Therefore, we can use the migrated tests for testing DL libraries into other domains of software.
% \victor{I did not get the ``therefore''}
% First, by means of \tool{}, we can test other frontends of DL compilers by migrating the test inputs for testing the corresponding DL libraries.
% Second, the migrated DL models can be directly leveraged to test other infrastructures of deep learning, as long as they take DL models as input.
% For example, model converters like MMdnn~\cite{mmdnn}, are designed to transform a DL model under one DL library into the equivalent DL model under another DL library.
% Since such converters also take DL models under various DL libraries as input, the migrated models can be used to validate their functionality.
\smallskip
\noindent\textbf{Improving regression testing.}
The migrated tests by \tech{} can help enrich regression test suites of DL compilers due to the diversity of bugs they detected.
There are 39 migrated tests by \tech{} that have been integrated into the official test suites of the DL compilers by developers, which have been used for regression testing in Continue Integration (CI).
Moreover, the test prioritization strategy in \tech{} has been demonstrated effective, which can be also used for optimizing the execution of regression tests in DL compilers.

\smallskip
\noindent\textbf{Coverage-based testing of the model loading stage.} Operator coverage plays a critical role in testing the model loading stage.
This suggests that if some automatic test generation techniques are designed, they can take operator coverage as guidance.
Similarly, if we incorporate more migration sources into \tech{} according to the conclusions of the complementary effectiveness among different migration sources, operator coverage can be used as the acceptance criterion. 

\smallskip
\noindent\textbf{Stage-specific testing.}
Although some DL model generation techniques (e.g., NNSmith and COMET) were proposed.
Their effectiveness is quite limited in testing the model loading stage.
Similarly, the test prioritization strategies widely-studied in general testing cannot effectively improve the test efficiency for our migrated tests.
In contrast, the design of \tech{} (including both test migration and prioritization components) considers the unique characteristics of the model loading stage, achieving promising effectiveness and efficiency.
This highlights the importance of stage-specific testing, which can be generalized to improve the testing of other stages.

\smallskip
\noindent\textbf{Comparing with direct model generation.}
Although designing a technique to generate single-operator models directly based on existing test generation fuzzers (e.g., DocTer and DeepREL) guided by the two diversity metrics (presented in Section~\ref{sec:tcp}) can be efficient, it has little influence on the overall DL compiler testing. This is because the most time-consuming step for \tech{} is test execution (including model compilation) rather than test generation, which accounts for over 95\% of the total time. Additionally, separating the test prioritization step in \tech{} helps optimize the execution of tests from multiple sources (including human-written tests and the tests generated by different fuzzers), indicating a global optimization strategy. In contrast, the test prioritization conducted by a fuzzer at each iteration is a local optimization strategy and it can not migrate tests from human-written tests in DL libraries, which actually detected the most DL compiler bugs (presented in Section~\ref{sec:result_sources}), for testing DL compilers. 
Hence, we proposed a migration-based testing technique (i.e., \tech{}) rather than designing a model generator directly, considering the generalizability and effectiveness.

\smallskip
\noindent\textbf{Avoiding false positives.}
Undefined behaviors and randomness are two main reasons leading to false positives during the testing process with migrated tests by \tech{}.
The method of avoiding false positives caused by randomness has been discussed in Section~\ref{sec:rq1_fp}.
However, there is still no method that can automatically detect undefined behaviors in operators, leaving the elimination of false positives caused by undefined behaviors as an open challenge.
Borrowing the knowledge in detecting traditional undefined behaviors~\cite{shen2021impact,lee2017taming} may help relieve this problem, which can be regarded as our future work.

\smallskip
\noindent\textbf{Community appreciation.}
Besides confirming and fixing our detected bugs and integrating some of our migrated tests into their official test suites, the TVM community also appreciated our contribution in the TVM forum many times, e.g., ``Detecting and fixing frontend bugs is very important work. You make a great contribution''.
In particular, the TVM community has invited the first author of this work to join them as a reviewer for TVM, because of the contribution of ``continuously improving frontend''.

\smallskip
\noindent\textbf{Threats to Validity.}
% \subsection{Threats to Validity}
The threats to validity mainly lie in the subjects and metrics.
To ensure generalizability of \tech{},
we considered eight frontends from the three popular DL compilers (i.e., TVM, TensorRT, and OpenVINO) for evaluation following the existing studies~\cite{ma2022hirfuzz,tzer}.
% To further reduce this threat,  DL compilers in total . 
% \scc{Discussion of the generalizability of the findings based on the tests generated by \tool~and the implemented prioritization strategy is not obvious.}
% To reduce the influence of randomness, we repeated our experiments five times and calculated the average results for illustration as discussed in Section~\ref{sec:process}.
There are also some metrics for evaluation test prioritization, such as APFDc~\cite{apfd} and RAUC-k~\cite{rauc,wang2021prioritizing}.
In our work, we used the most widely-used APFD metric and showed the trend of the number of detected bugs with the testing process proceeding.
Moreover, we also used the RAUC-k metric but put the results at our project homepage due to the space limit and consistent
% \jj{consistent?}\qingchao{yes}
conclusions.
% \qingchao{Done: human subjective threat, plz review!}

Besides, our work may suffer from the human subject threat as we reported the detected bugs to developers for confirmation and fixing.
To reduce this threat, we de-duplicated all test failures (see Section~\ref{sec:metrics})
and only reported the unique bugs.
All the responses from developers are positive and we received appreciation and confirmation from DL compiler communities as well, which reduces this threat.

% \subsection{Future Works}
% \victor{usually we don't have this section in submission}
% In this work, we only focus on test migration from DL libraries to DL compilers and have evaluated the effectiveness of detecting new bugs in DL compilers. 
% The test migration idea can be migrated for more testing scenarios.
% For example, migrating tests from one DL library to test other DL libraries. Different DL libraries have almost the same function but different grammar. Thus migrating the test access to different DL libraries is an interesting research direction.
% Also, migrating tests across different DL compiler frontends is a promising idea.

% For instance, ONNX has 20 different Opset versions, each including dozens of operator changes~\cite{onnx_opset}.
% TCP challenge: 
% \tool{} proposes an effective operators-based strategy.
% TODO: move the following to results analysis.
% Specifically, with the code change about operator implementation in DL frameworks, new test cases will be added in the built-in cases, which enriches the collected test cases.
% When the collected test cases update, test case prioritization needs to be executed repeatedly, which asks for high efficiency for the TCP algorithm.
% Therefore, \tool{} according to the test cases themselves rather than coverage information to rank them, which saves coverage collection time compared with coverage-based TCP algorithms~\cite{}.

%% file: 6-related.tex
\section{Related Work}
\label{sec:related}

% In this section, we introduce two lines of research 
% that are closely related to our study.
% \jj{please check whether there are new works on this topics. we should also discuss the works on test migration in general software testing as mentioned by R1.}

\subsection{DL Compiler Testing}
Recently, several techniques have been proposed for testing DL compilers.
According to the format of generated tests, they can be divided into IR-based test generation and model-based test generation.
The former directly skips the model-loading stage and targets the testing of compiler optimizations, including HirGen~\cite{ma2022hirfuzz}, Tzer~\cite{tzer}, and TVMFuzz~\cite{TVMFuzz}.
% For example, 
% HirGen designs a high-level IR generator to detect optimization bugs in the high-level optimization stage of TVM.
% Tzer and TVMFuzz detect bugs in the low-level optimization stage of TVM by mutating existing low-level IRs and generating low-level IRs from scratch, respectively.
% \victor{I did not get the meaning of ``separately'' here.}\qingchao{TVMFuzz uses the generation from scratch method, and Tzer use the mutation strategy.}
% \victor{should be respectively. Please comment this line after you read it}\qingchao{Yes, respectively rather than separately}
% All of these IR-style test generation techniques skip the model loading stage.
% Thus they are not able to detect the bugs in the model loading stage.
MT-DLComp~\cite{MT_DLComp} and NNSmith~\cite{NNSmith} are of another category, \ie, model-based test generation, which can cover all stages of DL compilers.
MT-DLComp proposes semantics-preserving mutation to generate equivalent DL models to support metamorphic testing.
NNSmith, the state-of-the-art technique, constructs DL models from scratch based on the corresponding grammar.
Both of them mainly focus on generating valid and diverse models to comprehensively trigger bugs in optimization stages. 
As they can only generate DL models represented under the ONNX library and support a limited number of operators, they are ineffective in testing the model loading stage.

% Different from them, our work explores the idea of migrating knowledge embedded in DL library tests for testing the model loading stage through an extensive study with \tech{}.
% This idea can obtain DL models represented under various DL libraries in a lightweight manner.
Different from them, the goal of \tool{} is to enhance the testing of the model loading stage in DL compilers.
Its core idea is to migrate test inputs from DL library testing, which can obtain DL models represented under various DL libraries in a lightweight way.
% In contrast, \tool{} mitigates the gap by migrating tests 
% from different sources in diverse DL libraries.
% Our experimental results 
% shows that \tool{} is effective in detecting bugs in the model loading stage.
% Almost all these detected bugs cannot be detected by the state-of-the-art technique (i.e., NNSmith).
% Moreover, NNSmith detects some optimization bugs that cannot be detected by \tool{}.
% The results 
% indicate that the bug detection abilities of \tool{} and NNSmith are orthogonal to some extent.
% Meanwhile, \tool{} can not detect most of the bugs found by NNSmith.
% Therefore, their bug detection ability is almost orthogonal.

\subsection{DL Library Testing}
Many techniques have been proposed to test DL libraries.
According to the test format during the generation of tests, they can be mainly divided into graph-level~\cite{lemon,comet,muffin,audee} and API-level~\cite{docter,deepREL,deltafuzz,Freefuzz,skipfuzz} test generation.
In the first category, CRADLE~\cite{cradle} makes the first attempt to test DL libraries with differential testing. 
Subsequently, LEMON~\cite{lemon}, Audee~\cite{audee}, EAGLE~\cite{eagle}, and COMET~\cite{comet} are proposed to generate DL models using a set of mutation rules.
% Inspired by EMI~\cite{emi}, EAGLE~\cite{eagle} designs several semantics-preserving mutation rules to enhance the test oracle ability.
% In addition, Muffin~\cite{muffin} and $\triangledown$Fuzz~\cite{deltafuzz} are proposed to detect bugs related to model training and automatic differentiation, respectively.
% \victor{Muinfin or Muffin?}
In the API-level test generation, 
Predoo~\cite{predoo} takes the first step to test DL libraries at the operator level.
It mutates the original tests to maximize output precision errors.
% FreeFuzz~\cite{Freefuzz} collected code snippets from the real world to test DL libraries.
TitanFuzz~\cite{TitanFuzz} utilizes LLMs to generate and mutate tests for testing DL libraries.
% The state-of-the-art technique, i.e., 
% DocTer~\cite{docter} generates test inputs based on the API constraints extracted from official documentation in DL libraries.

Unlike them, we proposed the idea of test migration from 
% of DL compiler testing 
DL library testing to enhance the testing of DL compilers.
% To test DL compilers, \tool{} migrates the tests from DL libraries, a nearby domain of DL compilers.
The tests generated by these DL library testing techniques can be the migration sources of \tool{}.
Indeed, \tool{} has integrated the tests generated by DocTer and DeepREL as migration sources.
% the state-of-the-art 
% DocTer as a migration source.
% in our work.
In the future, we can incorporate more techniques to enrich the migration sources of \tool{}.
That is, our methodology is orthogonal to DL library testing techniques.

% By contrast, existing techniques only consider the sources from the same domain of the software under testing.

% \subsection{Test Prioritization}
\subsection{Test Migration}
% \qingchao{this section needs to be reviewed}
Several test migration techniques~\cite{qin2019testmig,talebipour2021ui,behrang2019test,elbaum2008carving,zhong2022enriching,abdi2022test} have been proposed for various software.
For example, Sebastian et al.~\cite{elbaum2008carving} proposed a framework to extract differential unit tests from system tests for regression testing.
 % to test its subsequent software versions. 
% \jj{I cannot see it is relevant to test migration from the description.}.\qingchao{changed, plz review.}
Zhong et al.~\cite{zhong2022enriching} designed LERE to extract tests from bug reports of one traditional compiler
% (e.g., LLVM) 
to detect bugs in another compiler.
% (e.g., GCC).
% 
% Abdi et al.~\cite{abdi2022test} proposed a graph-based technique to extract tests
% \jj{what is it? hard to understand.} 
% from a dependent object-oriented project to test the imported libraries.
% Behrang et al.~\cite{behrang2019test} transformed the sequences of GUI events and oracles in the tests for the source mobile app to test the target mobile app that shares part of their functionality.

Different from them, \tool{} migrates knowledge from DL library testing to enhance the testing of DL compilers.
Our new scenario brings unique challenges for test migration, making the existing techniques inapplicable.
The significant challenge is to handle the fundamental difference between DL library testing and DL compiler testing.
As most tests for DL libraries do not include complete DL models but are just Python code, \tool{} instruments to extract DL operators and wraps them as single-operator models with templates to fill the gap.
Besides, a novel test prioritization strategy specific to our new scenario is designed for improving test efficiency, which has been demonstrated more effective than the existing test prioritization strategies for general software testing in our study.
% The scenario (i.e., across sibling tasks in the DL domain), technique challenge (i.e., huge difference between DL libraries testing and DL compiler testing), and method (i.e., code instrumentation and model wrapping) of migration in this work fundamentally differ from traditional methods.

%% file: 7-conclusion.tex
\section{Conclusion}
\label{conclusion}
% In this work, we conduct the first empirical study to investigate the effectiveness and efficiency of migrating the knowledge embedded in DL library tests to test the model loading stage of DL compilers.
% To enable the study, we develop \tool{} to extract operator instances from DL library tests and wrap them based on templates into DL models as migrated tests.
% To investigate efficiency improvement of such the migration-based idea, \tool{} includes a diversity-based test prioritization strategy.
% Based on three migration sources, we investigated the migration idea supported by \tech{} on eight frontends of three DL compilers.
% In total, these migrated tests detected \TotalBugsNum{} previously unknown bugs (including \ConfirmedBugsNum{} confirmed bugs). 
% The diversity-based test prioritization strategy in \tool{} achieves the average improvement of \APFDFastImprove{}$\sim$\APFDTotalImprove{} compared to general test prioritization in terms of APFD.
% \jj{wordy--changed}
% \qingchao{
In this work, we propose \tool{}, a migration-based technique, to test the model loading stage in a lightweight manner.
% (compared to grammar-based techniques with substantial implementation effort for supporting DL models under various DL libraries).
\tool{} uses tests documented in DL libraries, and the tests
generated by two recent fuzzers as migration sources.
Then, \tool{} extracts the operator instances from DL library tests and wraps them based on templates into DL models as migrated tests.
To improve the testing efficiency, 
\tool{} includes a diversity-based test prioritization strategy.
% \tool{} designs a diversity-based test prioritization strategy in order to detect more bugs within a given time budget.
By applying \tool{} to eight frontends of three popular DL compilers (i.e., TVM, TensorRT, and OpenVINO), \tool{} detected \TotalBugsNum{} previously unknown bugs (including \ConfirmedBugsNum{} confirmed bugs). 
The diversity-based test prioritization strategy in \tool{} achieves the average improvement of \APFDFastImprove{}$\sim$\APFDTotalImprove{} compared to general test prioritization in terms of APFD.

% \section{Data Availability}
% \label{data}
% To facilitate future work and experimental replication, we released the artifact of \tool{} (including code, datasets, and other supplementary materials) at the project homepage: \url{https://github.com/AnonymousWorks/OPERA}. 

%% file: 8acknowledge.tex
\section*{Acknowledgments}
\label{sec:acknowledge}
We thank all the anonymous reviewers for their thoughtful and constructive comments on this work.
This work was supported by the National Natural Science Foundation of China (Grant Nos. 62322208, 12411530122, 62232001), CCF Young Elite Scientists Sponsorship Program (by CAST), and Hong Kong Research Grant Council/General Research Fund (Grant No. 16205722).